\documentclass[twocolumn]{aastex62}

\usepackage{amsmath}
\usepackage{natbib}
\usepackage{graphicx}
\usepackage{epstopdf,color}
\usepackage{verbatim}
\usepackage{comment}
\usepackage{hyperref}
\bibpunct[]{(}{)}{;}{a}{}{;}

\newcommand\eris{\emph{Eris}}
\newcommand\eiik{\emph{Eris2k}}
\newcommand\erisnfb{\emph{ErisNFB}}
\newcommand\venus{\emph{Venus}}

\newcommand{\arif}[1]{{\leavevmode\color{black}#1}}

\newcommand{\arifho}[1]{{\leavevmode\color{black}#1}}

\received{}
\revised{\today}
\accepted{}
\submitjournal{ApJ}

\shorttitle{Gaseous Galactic Halos}
\shortauthors{Soko{\l}owska et al.}

\begin{document}

% LaTeX will automatically break titles if they run longer than
%% you desire. In v6.2 you can include a footnote in the title.

\title{The complementary roles of feedback and mergers in building \\ the gaseous halo and the X-ray corona of Milky Way-sized galaxies}

\correspondingauthor{A. Babul}
\email{babul@uvic.ca}

%% A significant change from earlier AASTEX versions is in the structure for calling author and affilations. 
%%
%% The \author command is the same as before except it now takes an optional argument which is the 16 digit ORCID. The syntax is:
%% \author[xxxx-xxxx-xxxx-xxxx]{Author Name}
%%
%% This will hyperlink the author name to the author's ORCID page. 
%%
%% Use \affiliation for affiliation information. The old \affil is now aliased  to \affiliation. AASTeX v6.2 will automatically index these in the header.  When a duplicate is found its index will be the same as its previous entry.
%%
%% Please use multiple \affiliation calls for to document more than one affiliation.
%%
%% The new \altaffiliation can be used to indicate some secondary information such as fellowships. This command produces a non-numeric footnote that is set away from the numeric \affiliation footnotes.  NOTE that if an \altaffiliation command is used it must come BEFORE the \affiliation call, right after the \author command, in order to place the footnotes in the proper location.
%%
%% Use \email to set provide email addresses. Each \email will appear on its own line so you can put multiple email address in one \email call. 
%%

\author{A. Soko{\l}owska}
\affil{Center for Theoretical Astrophysics and Cosmology, University of Zurich, Winterthurerstrasse 190, Zurich, Switzerland.}

\author{A. Babul}
\affil{Center for Theoretical Astrophysics and Cosmology, University of Zurich, Winterthurerstrasse 190, Zurich, Switzerland.}
\affil{Department of Physics and Astronomy, University of Victoria, Elliot Bldg, 3800 Finnerty Rd, Victoria, Canada.}

\author{L. Mayer}
\affil{Center for Theoretical Astrophysics and Cosmology, University of Zurich, Winterthurerstrasse 190, Zurich, Switzerland.}

\author{S. Shen}
\affil{Kavli Institute for Cosmology, University of Cambridge, Madingley Road, Cambridge CB3 0HA, UK.}
\affil{Institute of Theoretical Astrophysics, University of Oslo, Sem Saelands vei 13, Svein Rosselands hus, 0371 Oslo, Norway.}

\author{P. Madau}
\affil{Center for Theoretical Astrophysics and Cosmology, University of Zurich, Winterthurerstrasse 190, Zurich, Switzerland.}
\affil{Department of Astronomy and Astrophysics, 1156 High Street, University of California, Santa Cruz CA 95064, USA.}

%%%%%%%%%
%ABSTRACT
%%%%%%%%%
\begin{abstract}
\arif{
We use high-resolution cosmological hydrodynamical simulations of Milky Way-sized galaxies with varying supernovae feedback strengths and merger histories to investigate the formation of \arif{their gaseous halos and especially their  hot ($>10^6$~K) X-ray luminous coronae.  Our simulations predict the presence of significant hot gas in the halos as early as $z=3-4$, well before the halos ought to be able to sustain hot mode accretion in the conventional picture. The nascent} coronae grow inside-out and initially do so \arif{primarily as a result of outflows from the central galaxies powered by} merger-induced shock heating and strong supernovae feedback, both of which are elemental features of today's successful galaxy formation models.  Furthermore, the outflows and the forming coronae also accelerate the transition from cold to hot mode accretion by contributing to the conditions for sustaining stable accretion shocks.  They also disrupt the \arif{filamentary streams funneling cold gas onto the central galaxies by causing their mouths to fray into a broad delta, detach from the galaxies, and be pushed away to larger radii.  And even though at early times the filaments repeatedly re-form, the hot gas and the outflows act to weaken the filaments and accelerate their ultimate disruption.  Although galactic outflows are generally thought of as ejective feedback, we find that their action on the filaments suggests a preventive role as well.  
}\\

%From the time this starts to when the halos fully transition to hot mode accretion, the filaments repeatedly re-form and fray until they finally cease to exist. 
%We find that the mergers and strong feedback, , alter for the evolution of the gaseous halos with respect to the conventional picture, in which the the halo gas either enters the halo via the filaments and heats up due to compression, or accretes onto the halo quasi-spherically and is heated by an accretion shock at the virial radius.
}
%All this is quite different from the standard picture in which diffuse halos are \textbf{only} a consequence of the thermalisation of kinetic energy derived from gravity and/or the geometric effect of cross sections of halos vs. filaments, and may be more relevant for halos harbouring typical spiral galaxies. We show that SN feedback impacts the galaxy-cold flows connection, which has also consequences for the large-scale gas supply and may contribute to galaxy quenching.
\end{abstract}
\section{Introduction}
\label{sec:intro}

%\arifhum{NEED TO GO BACK OVER THIS AND REVIEW VOORT ET AL AND COOLING ISSUES}

\arif{Studying the formation and evolution of Milky Way (MW)-like galaxies with stellar masses ranging from $3\times 10^{10}\;{\rm M_\odot}$ to $3\times 10^{11}\;{\rm M_\odot}$ has been a focus of concerted attention for several decades.  These systems 
are ubiquitous --- they comprise the majority of the bright galaxies in the local universe --- and they host more than two-thirds of the present-day stellar-mass density (see \citealt{Papovich:2015} and references therein).   

Although the majority of studies focus on formation and evolution of the central galaxies themselves \cite[see, for example,][]{Guedes:2011aa,Stewart:2017,GK:2017}, one of the earliest predictions of the hierarchical structure formation models is that these galaxies ought to be cocooned by halo-filling diffuse gas, with temperatures ranging from a few $\times 10^5$ to a few $\times 10^6$~K \citep{ReesOstriker:1977, WhiteRees:1978}.  

Until about a decade ago, observational evidence for the existence of this gaseous halo was sparse and difficult to interpret \citep[see the review on gaseous halos by ][]{Putman:2012aa}.  However, significant recent progress has resulted in a compelling compendium of UV and X--ray absorption line data \citep{Tumlinson:2011, Peeples:2014, Gupta:2012,Miller:2013aa, Fang:2015aa}, OVII and OVIII emission line measurements \citep{Gupta:2009, Henley:2012a, Henley:2013aa, Miller:2014aa}, and even the direct detection of extended X-ray emission from bright spiral galaxies more massive than the MW \citep{Anderson:2011aa, Bogdan:2012aa, Bogdan:2013b,Anderson:2016}. 
%This circum-galactic media have already been detected in ultraviolet and X--ray absorption lines \citep{Sembach:aa, Wang:2005,Bregman:2007aa, Peeples:2014, Miller:2014aa, Fang:2015aa}, in OVIII emission \citep{Gupta:2009, Henley:2013aa}, and as excess X--ray emission relative to the background \citep{Rasmussen:2009aa,  Anderson:2011aa, Bogdan:2013b, Tumlinson:2011}. 
%The direct detection of hot halo gas ($T>10^6$~K) at extragalactic distances is currently possible only around massive galaxies \citep[e.g.][]{Anderson:2011aa,Bogdan:2012aa}, but in the future a more sensitive X-ray observatory such as \emph{Athena+} may bring spatially resolved images also of lower-mass spirals \citep{Kaastra:2013}. 
%For galaxies with stellar masses $\lesssim 5\times 10^{10}\; M_{\odot}$ and $M_{halo} \lesssim 10^{12} M_{\odot}$, however, our 
Nonetheless, our Galaxy is presently the only available laboratory for studying in detail the structure and the characteristics of the gaseous halos around ordinary MW-like galaxies, and potentially gain insights into their origin.
%For the time being, the hot halo around our Galaxy is the best probe of hot halos around lower-mass galaxies ($M_{halo} \simeq 10^{12} M_{\odot}$). Recent modeling of a million degree gaseous halo in the Milky Way results in a hot halo mass estimate of $M(200 kpc) = 3.8 \times 10^{10} M_{\odot}$  \citep{Miller:2013aa}. 
The latest observational analysis indicates that the total mass of this component around the Milky Way is $\simeq 3-5\times 10^{10}\; M_{\odot}$ and extends essentially out to the virial radius  \citep{Miller:2014aa}, in excellent agreement with the results from the \eris~simulations discussed in \citet[][]{Sokolowska:2016}.

In existing structure formation models, the origin of the gaseous halo is intimately linked to the evolution of the dark matter halos and their central galaxies.  In the earlier incarnations of the galaxy formation picture, gas flowing onto a growing dark matter halo shock-heats to form a diffuse, dark matter halo-filling atmosphere. This gaseous halo was assumed to be quasi-static if its ``cooling radius'' --- the radius where the characteristic cooling time equals the age of the universe --- is close to the halo center.  Otherwise, the gas within the cooling radius cools, sinks towards the halo center, and fuels both the formation of, and the star formation within, the central galaxy \citep{ReesOstriker:1977,Binney:1977,Silk:1977,WhiteRees:1978,WhiteFrenk:1991}.  

A complication arises when the cooling radius exceeds the virial radius.  As discussed by \citet{WhiteFrenk:1991}, cooling is so rapid that the infalling gas never establishes an extended quasi-static atmosphere.  Although not explicitly stated at the time, this also means that a stable accretion shock cannot form at the virial radius, and the infalling gas will flow onto the galaxy without experiencing an accretion shock.
%In these models, whether the accretion shock develops at the virial radius or closer to the central galaxy and whether a diffuse gaseous halo emerges or not depends on the ratios of the cooling and the Hubble times, and the cooling and the gravitational collapse times \citep{ReesOstriker:1977,WhiteFrenk:1991}.
%The question of the origin of diffuse gas has been under investigation for several decades. Until about a decade ago, the standard paradigm of galaxy formation assumed that dark matter relaxes to a virial equilibrium, and gas follows the dark matter during collapse \citep{ReesOstriker:1977, WhiteRees:1978}. The resulting supersonic accretion triggers an accretion shock \citep{Binney:1977}. The shock develops at the virial radius or closer to the galaxy, depending on the ratio of the cooling and dynamical timescales \citep{WhiteFrenk:1991}. 
%halos in the mass range $M_{halo} \simeq 10^{12-13} M_{\odot}$ are expected to support an accretion shock at their virial radius, and thus contain a quasi-hydrostatic atmosphere of hot gas.  A more detailed assessment of shock stability by \citet{Birnboim:2003}
Based on their shock stability analysis, \citet{Birnboim:2003} identified the critical halo mass at which a stable shock can be maintained at the virial radius to be $\simeq 10^{11} - 10^{12} M_{\odot}$, depending on the metallicity of the gas.   
%A detailed analytical treatment of shock stability in \citet{Birnboim:2003} brought down the critical halo mass for a stable shock at the virial radius to over $10^{11} M_{\odot}$ for primordial gas and around $10^{12} M_{\odot}$ for solar metallicity gas. The accretion shock relies on the presence of a stable atmosphere of post-shock gas to support itself. 

The above picture, and the calculations, are, however, rooted in the idealized, spherical collapse and spherical halo models.  In practice, both the accretion history and the flow geometry of the mass converging onto a halo is much more complex.  In lock-step with the emergence of virialized structure, the large-scale matter distribution in the universe organizes into a network of filaments and sheets; and as the virialized structures grow hierarchically, the filaments too merge and grow in scale \citep{CosmicWeb:2014}.  Initially, the halos of MW-like systems form at the intersections of filaments but eventually, they are incorporated into either larger filaments (field galaxies) or larger virialized systems (group and cluster galaxies).
%However, both the models of \citet{WhiteFrenk:1991, Birnboim:2003} assumed spherical symmetry, and thus could not capture the scenario of asymmetric halo configurations. 

Successful execution of increasingly realistic, high resolution galaxy formation simulations within their proper cosmological setting  has radically transformed our understanding of how baryons accrete onto galactic halos and how galaxy formation proceeds. For one, the bulk of the gas accreting onto a halo at a redshift $0 \leq z \lesssim 3$ is never heated to the halo's virial temperature if the halo mass is less than $3\times 10^{11} M_{\odot}$ \citep{Fardal:2001, Katz:2003, Keres:2005, Keres:2009, Dekel:2009, Brooks:2009}.  
%3D simulations have revealed that most of the gravitational cooling radiation comes from gas at temperatures far below the typical virial temperatures of galaxies \citep{Fardal:2001},and that a significant fraction of gas in galaxies at the low-mass end of this range has never been shock heated \citep{Katz:2003, Keres:2005, Keres:2009, Dekel:2009, Brooks:2009}. \citet{Keres:2005, Keres:2009} showed that, for all halos below a critical mass of $3\times 10^{11} M_{\odot}$ and at redshifts in range $z=0-3$, the infall is predominantly cold. 
In fact, the simulations suggest two distinct modes of smooth gas accretion {\it onto the galactic halos}: (i) ``cold mode" accretion, where the gas accretes via ``cold'' filamentary streams that often penetrate to the halo centers; and (ii) ``hot mode" accretion, where the accreting gas encounters a stable accretion shock and the combination of this and the subsequent compressive heating raises its temperature to 
$T\sim {\rm a\; few}\; \times 10^5$ to as much as ${\rm a\; few}\; \times 10^6$~K  
\citep[see][]{Keres:2005, Keres:2009, Dekel:2006, Dekel:2009}.
%(filamentary inflow resulting in the maximum gas temperature that is lower than the virial temperature) and ``hot mode" \citep[spherical accretion with heating of gas up to $T>10^{5-7}$~K before cooling, see also][]{Dekel:2006, Keres:2009, Dekel:2009}. 

In this revised paradigm, initially the gas enters the galactic halos
through narrow, cold, dense filamentary streams.  As the streams converge onto the central galaxy, they experience compressive heating but while the gas cooling time is shorter than the heating time, the streams remain cold and deposit their gas directly onto the central galaxy \citep{Voort:2011,Stewart:2017}.  Eventually, the cooling efficiency of the gas drops, the cooling time exceeds the compressive heating time, and an increasing fraction of the gas  starts to form the galaxy's diffuse gaseous halo instead of settling into the galaxy  \citep{Joung:2012}.  The emergence of this atmosphere sets the stage for a stable accretion shock at the virial radius. 
%once halo mass exceeds the critical mass threshold mentioned above.  

Concurrently, on the supra-galactic scale, the cosmic filaments are growing and as their widths become comparable to the sizes of the halos, an increasing fraction of the gas flows onto the halos in quasi-spherical fashion, shock heating as it encounters the accretion shock, and then heating up further as it is squeezed deeper into the halo.  It is this gas that is generally thought to build the diffuse gaseous halos \citep{Voort:2011}.  

Before proceeding, we emphasize that most of the discussion of hot and cold mode accretion in the literature refers to the accretion onto the central galaxy.  Here, we are mainly interested in gas accretion onto the galactic halos\footnote{For a thorough discussion of the similarities and difference between gas flow onto the halo and onto the galaxy, we refer the reader to \citet{Voort:2011}.} as it pertains to the establishment of the gaseous halo
and especially, the X-ray  luminous corona. We also caution that the simple picture sketched out here should be viewed as a conceptual aide.  In practice, the different stages do not unfold sequentially and there are times when both the hot and the cold modes co-exist.  
In general, the timing of the transition from predominantly cold to largely hot mode accretion depends on each halo's individual growth rate but 
for MW-like galaxies, it typically occurs around $z\sim 2$ \citep{Keres:2009}.
%The diffuse X-ray luminous halo is thus generated by the hot mode accretion, in halos which grow sufficiently in mass (i.e. beyond the critical mass threshold). The time of this transition depends on the halo growth rate but is often seen around $z\sim 2$. 

The studies of \citet[][]{Fardal:2001, Katz:2003, Keres:2005, Keres:2009, Dekel:2006} that laid the foundation for the above picture, though groundbreaking, did not include the full set of sub-grid implementations of the key physical processes that gird today's successful galaxy formation simulations.  The simulations of \citet{Keres:2005, Keres:2009}, for example, lack sufficiently effective supernovae (SNe) feedback required to prevent the over-production of stars.  Contemporary studies have shown that strong SNe feedback, in concert with other stellar feedback processes like stellar winds, radiation pressure, and photo-heating, is crucial for (i) obtaining realistic stellar and baryonic masses and star formation efficiencies in halos with masses as high as $10^{13} M_{\odot}$ \citep{Dave:2011,Dave:2012,In-N-OutBaryons:2016,Liang:2016}; (ii) ensuring that the lower-mass end of the galaxy stellar mass function agrees with observations
\citep[e.g.][]{Dave:2011,Bower:2012,Puchwein:2013}; and (iii) realizing disk galaxies with reasonable global and local properties across a broad mass range: dwarfs \citep{Shen:2014}, spiral galaxies \citep{Guedes:2011aa,Marinacci:2014,Agertz:2014,Roskar:2014,GK:2017}, and massive early--types \citep{Argo:2014}.  

A common feature of all the simulations is that strong SNe feedback engenders powerful galactic outflows \citep[see the review by][ and references therein]{SomervilleDave:2015}. This is not a surprise; numerous studies dating back to \citet{Mathews:1971,Larson:1974,DekelSilk:1986,BabulRees:1992} have anticipated this.  Moreover, it seems inevitable that the expulsion of a large fraction of the gas out of the galaxies and into the halo (and beyond) ought to have a profound impact the emergence and evolution of the gaseous halo.  This was hinted at by \cite{WhiteFrenk:1991}, and  several simulation-based studies \citep[e.g.~][]{
Oppenheimer:2006,Dave:2011,Dave:2012,Voort:2011,Voort:2016,
In-N-OutBaryons:2016,Stewart:2017} of galaxy formation have also commented on this, noting that galactic outflows are required to account for the metallicity gradients within the gaseous halos; that they tend to make the gaseous halos more radially extended --- in the \eris~simulation of the MW galaxy, the total baryonic mass fraction within the virial radius is only 
$70\%$ of the universal value \citep{Sokolowska:2016} ---
%that the gas can, at times, be flowing outward at velocities close to the systems' escape velocity; 
and that both the size and the mass of these halos are correlated with the star formation episodes \citep{Voort:2016}.
%and that strong outflows can temporarily disrupt the inflowing filamentary streams.   

In this paper, we present a detailed investigation of the origin of the gaseous halo around MW-like galaxies.  We pay particular attention to the %phenomena described above, as well as on the 
roles of the different accretion modes, merger-induced shocks \citep[see also][]{Sihna:2009}, and SNe feedback in building the gaseous halo.  The present study is 
motivated by our previous paper  \citep[][, hereafter Paper~I]{Sokolowska:2016}, where we explored the present-day diffuse halos of MW-like galaxies using a suite of high-resolution, zoomed-in hydrodynamic simulations.  Our principal simulation, \eris, employs a SNe feedback prescription that is remarkably successful at making a realistic Milky Way-like galaxy \citep{Guedes:2011aa}.  In Paper I, we showed that the key properties of this virtual MW's gaseous halo are also in excellent agreement with recent observational constraints for our Galaxy's halo: namely, an X-ray luminosity in the 0.5-2~keV band of $\sim 10^{39}$~erg/s; coronal density sufficient to ram-pressure strip Milky Way satellites; and a radial electron density profile consistent with that implied by the OVII absorption line measurements of \citet{Miller:2013aa} \citep[see][ for details]{Sokolowska:2016}.
 We also identified an intriguing feature in the halo of our simulated galaxy: a central spheroidal region with a radial extent of 100-140~kpc ($0.6~r_{vir}$) embedded in a $T=10^{5-6}$~K warm-hot atmosphere containing most of the mass.  Within this central region (hereafter, referred to as the corona), the density of the hot ($T>10^6$~K) gas exceeds that of the warm-hot gas.
 
%that we referred to as a \emph{corona}. The Coronae are central regions of hot gas with temperatures exeeding $10^6$~K, characterised by approximately spherical geometry and a high degree of localization, as they tend to extend no farther than . They are 

Here, we not only examine the build-up of the corona and the gaseous halo in \eris~ but also, for comparison, 
\arif{
in (i) \erisnfb, i.e.~\eris~with SNe feedback switched off, (ii) \eiik~\citep{Sokolowska:2016,Sokolowska:2017}, a variant with the same initial conditions as \eris~but with enhanced cooling as well as boosted SNe feedback,
\citep[see also][]{Shen:2012,Shen:2013} } and (iii) \venus~\citep{Sokolowska:2017}, 
a new addition to our inventory of zoomed-in simulations of MW-like galaxies with a more active merger history than \eris. 
}

The present paper is organized as follows: In Section 2, we motivate why we investigate this particular set of runs and describe the physics included in the simulations. In Section 3, we present our results about the emergence of the gaseous halos, including the inside-out growth of the hot corona, in Eris and Venus simulations.  We discuss the roles of stellar feedback and merger-induced shocks in giving rise to the structure seen in our simulations, and explore the connection between the growth of the corona and the inflowing cold filamentary network.  
%and the role of feedback in their formation. We also explore the connection between the growth of coronae and the cold filamentary network.  
This is then followed by a summary and concluding remarks in Section 4.  

\section{Simulations}\label{sec:simulations}
%Run, $z_l$, M$_{vir}(z_l)$, R$_{vir}(z_l)$, IMF, MC, UVB, $c^*$, $n_{SF}$, $\epsilon_{SN}$
%\eris, 0, 7.6, 233, K1993, low--T, HM1993, 0.1, 5, 0.8
%ELE, 0, 7.8, 235, K1993, low--T,HM1993, 0.05, 5, 0.8
%\eiik, 0.5, 6.5, 170, K2001,all--T,HM2012, 0.1, 100, 1.0

\begin{figure*}
\centering
\includegraphics[scale=0.5]{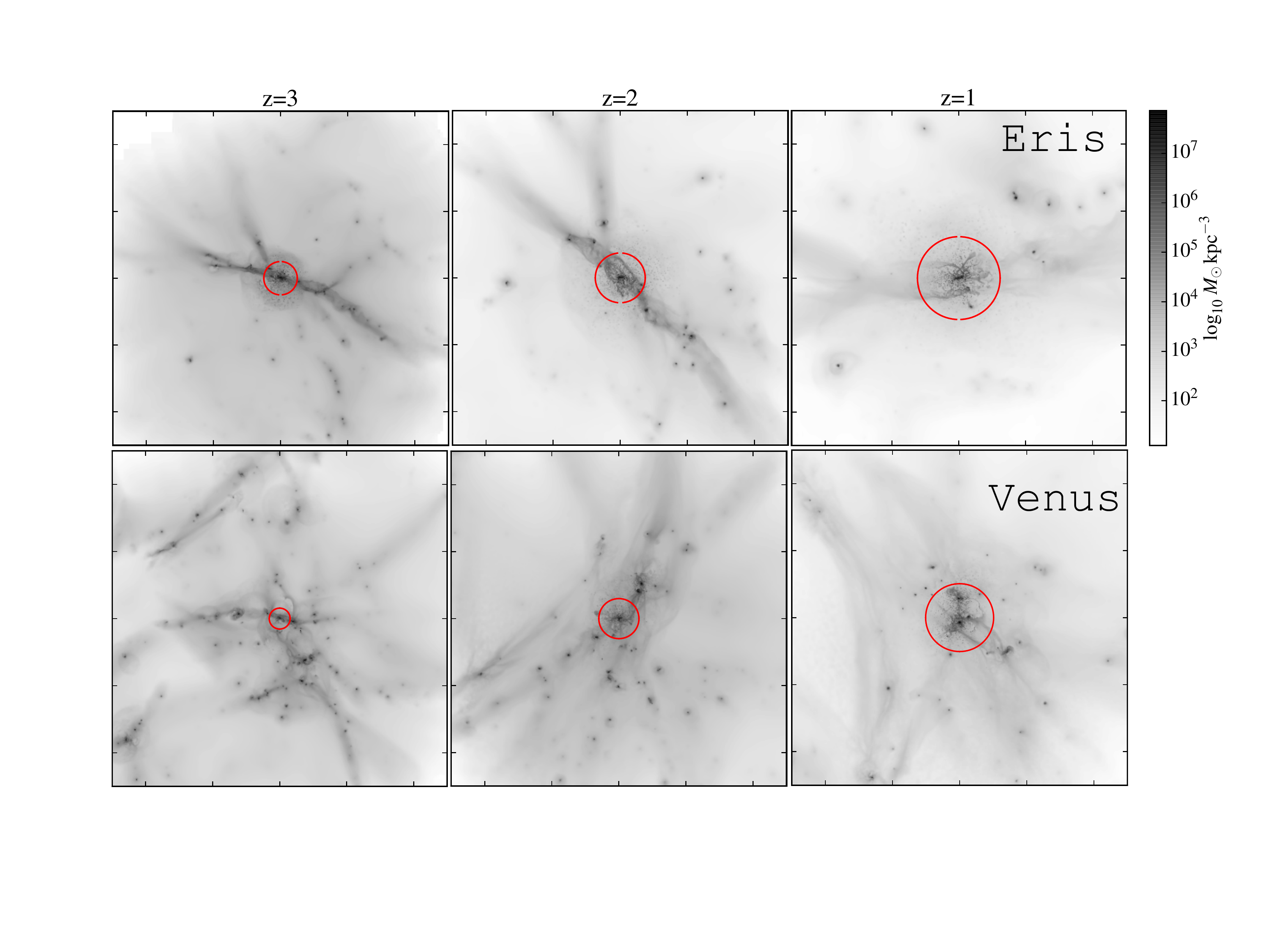}
\caption{Density maps of gas around the galaxies (shown edge-on) in the two simulations under consideration: \eris~ (quiet accretion history) and \venus~(active accretion history). Width of each square is 1 comoving Mpc. The cross sections of the most massive progenitor halo at each time step are shown as red circles.}
\label{fig:geometry}
\end{figure*}

Our study is based firstly on two unique, high-resolution, cosmological, zoom-in simulations of  Milky Way-sized spiral galaxies, \eris~ and \venus.  The simulations assume the flat WMAP-3 cosmology ($\Omega_M=0.24$, $\Omega_b=0.042$, $H_O=73$~km/s/Mpc, $n=0.96$, $\sigma_8=0.76$) and were performed with the tree-smoothed particle hydrodynamics (SPH) code {\sc gasoline} \citep{Wadsley:aa} with mass resolution $m_{\rm dm} \simeq 9.8 \times 10^4$~M$_{\odot}$ and $m_{\rm SPH} \simeq 2 \times 10^4$~M$_{\odot}$, and spatial resolution $\simeq 120$~pc.   

The late-time spiral galaxy in the first of the two runs, \eris, has been previously shown to be an excellent match to the Milky Way 
\citep{Guedes:2011aa}.  The second simulation, \venus, was run with the same sub-grid prescriptions as \eris~ but employs different initial conditions (IC) while ensuring that the final virial halo mass ($\sim 8 \times 10^{11}$~M$_{\odot}$) and the halo spin parameter ($\lambda \sim 0.03$) are nearly identical to that in \eris. 
%The ICs of \eris were generated with {\sc grafic2} \citep[][]{Bertschinger_2001}. The ``zoom-in'' of \venus was initialized using the {\sc music} code \citep{MUSIC}, which allows a computationally more efficient topological identification of the Lagrangian subvolume for the refinement. 
As a result, although the galaxies in \eris~ and \venus~ both form at the intersection of four dark matter filaments, their mass convergence pattern is very different (see Figure~\ref{fig:geometry}).  The dominant halo in \eris~assembles early on and has a relatively quiet merging history, with the last major\footnote{Defined as a merger with mass ratio $>0.1$ between the two galaxies.} merger occurring $z = 3.1$.  On the other hand, \venus~
has a much more active merger history and experiences twice as many major mergers as \eris, with the last major merger occurring
at $z = 0.9$. Also,  multiple progenitors of comparable mass evolve separately in \venus~for a long time, with single dominant halo and its associated galaxy only appearing after the last major merger. The amount of substructure at $z=0$ is also more abundant in \venus~ than in \eris, both in the stellar and in the dark matter components.  In particular, a large satellite orbits around the primary galaxy in \venus~ at late times, exciting perturbations in the disk of the central galaxy at pericenter passages as late as $z=0.24$.

%These ICs were chosen to have an active merging history down to a low redshift in contrast to the quiet merging history of \eris, 

As for the physical processes, both simulations include radiative and Compton cooling. The radiative cooling is computed for a simple mixture of H and He via non-equilibrium cooling rates in the presence of the ionizing cosmic ultraviolet (UV; \citealt{Haardt:1996}) background \citep{Wadsley:aa}. Additionally, $T<10^4$~K gas can cool through fine structure and metastable lines of C, N, O, Fe, S, and Si \citep{Bromm:2001aa, Mashchenko:2007aa}.  

Both runs utilize the same star formation prescriptions \citep[see][ for details]{Stinson:2006aa}.  Briefly, the gas particles must have a density greater than the star formation threshold of $n_{\rm SF} = 5$ atoms~cm$^{-3}$ and temperature  $T < 10^4$~K in order to form stars.  Particles that satisfy these requirements are stochastically selected to form stars according to $dM_*/dt = c^* M_{\rm gas}/t_{\rm dyn}$, where $M_*$ is the mass of stars created, $c^*$ is a constant star formation efficiency factor (set to 0.1 in all runs), $M_{\rm gas}$ is the mass of gas creating the star, and $t_{\rm dyn}$ is the gas dynamical time. Each star particle then represents a population of stars, covering the entire initial mass function (IMF). We adopt the \citet{Kroupa:1993aa} IMF and stars more massive than 8~M$_{\odot}$ are assumed to explode as SNII at the end of their lives.

In both runs, SNI and SNII enrich the interstellar medium (ISM) with metals as well as inject energy into the medium.  According to the ``blastwave feedback'' model of \citet{Stinson:2006aa}, SNe energy is introduced as purely thermal injection of energy as the blastwave shocks are expected to convert the kinetic energy of ejecta into thermal energy on scales smaller than those resolved by our simulations.  Once energy is introduced (the fraction of SN energy that couples to the interstellar medium is $\epsilon_{\rm SN} = 0.8$), the particles receiving the energy are prevented from cooling for typically 10--50~Myr, with the cooling shut-off timescale being computed as the sum of the Sedov--Taylor \citep{Taylor:1950,Sedov:1959} and snow-plough phases in the ejecta \citep{McKee_Ostriker_1977}. By delaying the cooling, we model, in a phenomenological way, the unresolved effects of momentum driving and heating by turbulent dissipation in the ejecta before they reach the radiative phase. Delaying the cooling also prevents artificial overcooling of gas heated by SNe feedback. The strength of feedback depends on the number of SNe produced, which is in turn governed by the IMF and, locally, by the star formation density threshold. 

Apart from \eris~and \venus, we also consider two additional simulations, both variants of \eris: \erisnfb~and \eiik.  \arif{Both runs have identical initial conditions as \eris.  \erisnfb~has the same  physics as \eris~except that the SNe do not inject any energy into the ISM (see Section 3.3) while in \eiik~(see Section 3.5), the physics set is augmented to allow gas at $T > 10^4$ K to cool via metal-line radiation, more efficient metal-line cooling of gas at $T < 10^4$ K \citep[c.f.][]{Shen:2012,Shen:2013}, SNe feedback is boosted, and turbulent diffusion of thermal energy and metals is enabled following Shen et al. (2010).  Metal-line cooling is computed with the code {\sc cloudy} \citep{cloudy} under the assumption that the metals are in ionization equilibrium \citep{Shen:2009aa}, in presence of an updated cosmic ionizing background \citep{Haardt:2012aa}. SN Feedback is enhanced both locally and globally by simultaneously raising the star formation density threshold to 100 atoms~cm$^{-3}$ (see \citealt{Guedes:2011aa,Mayer:2012,Hopkins_et_al_2012} for a discussion about the resulting local enhancement of SN feedback) and adopting a more top-heavy \citet{Kroupa:2001aa} IMF than the \citet{Kroupa:1993aa} IMF used in \eris, which yields about a factor of 2.8 more SNe at the same star formation rate.   We refer interested readers to \citep{Sokolowska:2016} and \citep{Sokolowska:2017} for a more details about the \eiik~simulation. Table~\ref{tab:tab0} summarizes the runs discussed in this paper.}

\begin{table}
\centering
\begin{tabular}{c|c|c|c|c|c|c|c}
Run & UVB & IMF & $n_{\rm SF}$ & $T_{\rm max}$ & $\epsilon_{\rm SN}$ & MC & IC\\
\hline
\eris		& HM96 & K93 & 5	& $3\times 10^4$ & 0.8 & low-T &  Q\\
\eiik		& HM12 & K01 & 100	& $1\times 10^4$ & 1.0 & all-T   & Q\\
\venus	& HM96 & K93 & 5		& $3\times 10^4$ & 0.8 & low-T &  A\\
\end{tabular}
\vspace{10.0pt}
\caption{Input parameters of the runs. Notation: UVB -- UV background (HM96: \citealt{Haardt:1996}, HM12: \citealt{Haardt:2012aa}), IMF -- initial mass function (K93: \citealt{Kroupa:1993aa}, K01: \citealt{Kroupa:2001aa}), $n_{\rm SF}$ -- star formation density threshold in $\rm cm^{-3}$, $T_{\rm max}$ -- maximum temperature (in K) of a gas particles that can participate in star formation,
$\epsilon_{\rm SN}$ -- SN efficiency parameter in $10^{51}$ erg, MC -- metal cooling, and IC -- initial conditions (Q: quiet merger history, A: active merger history).}
\label{tab:tab0}
\end{table}

%{Additionally, we investigate the consequences of the exclusion of feedback (model of \eris without supernova feedback, see section 3.2), as well as include a discussion of the impact of enhanced feedback on our results (model \eiik, see the Appendix).}

\begin{figure*}
\centering
\includegraphics[scale=0.8]{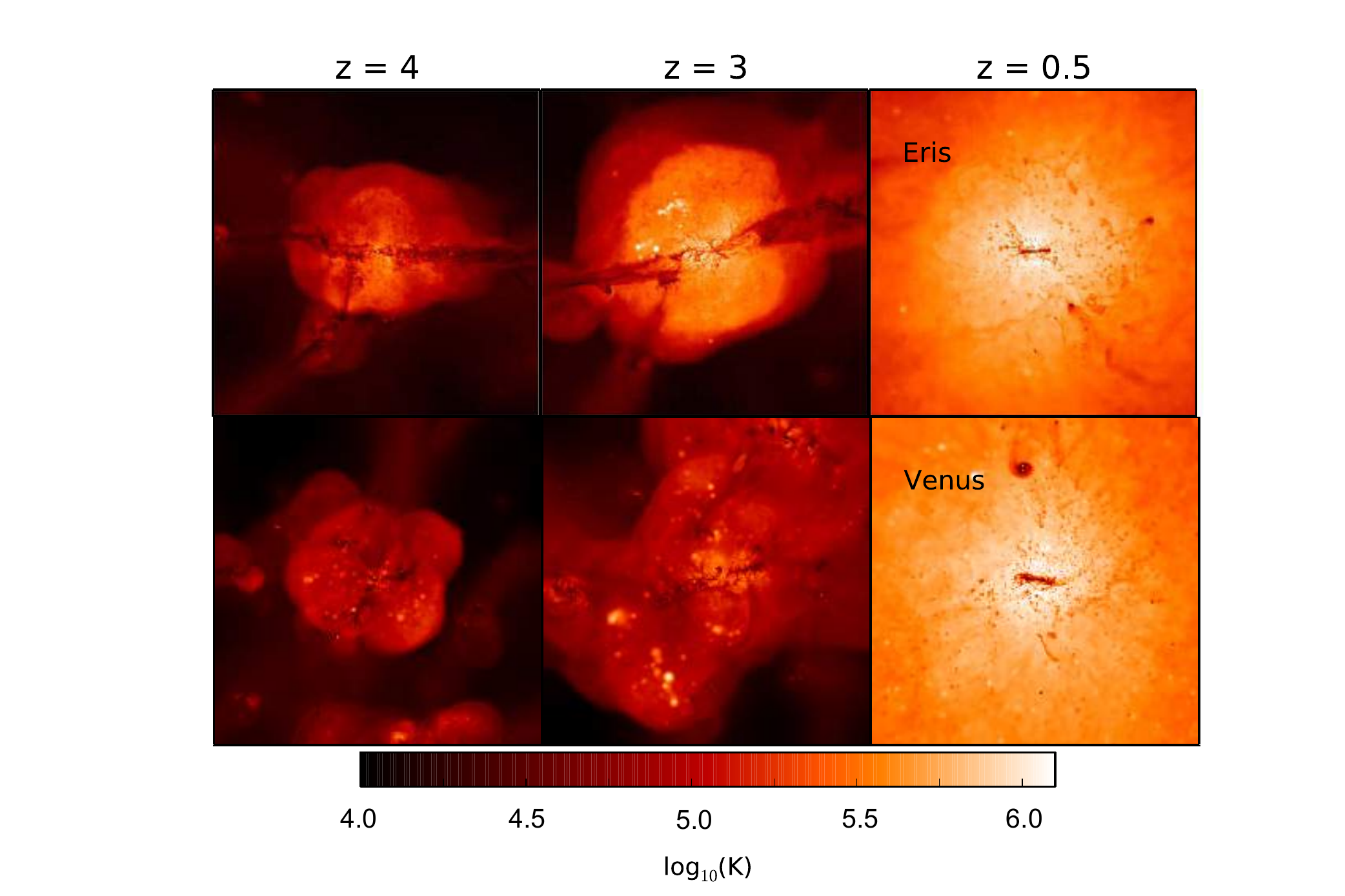}

\caption{ {Two pathways to the same result: a galaxy surrounded by an X-ray-bright corona.} Temperature maps of the gaseous halo of \venus~ and \eris~centered on the most massive progenitor halo, at three different times. The width of each square is 300~comoving kpc. Compare with Fig.~1 of \cite{Sokolowska:2016}. }
\label{fig:temp_map}
\end{figure*}

\section{Results}\label{sec:results}
{In Paper~I, we discussed in detail the properties and the structure of the low redshift gaseous halo in the \eris~simulation.   We found that this halo consists of a central spherical region about the primary galaxy, containing predominantly hot ($T>10^6$~K), and surrounded by an atmosphere for warm-hot ($10^{5-6}$~K) gas extending out to the virial radius.  We labeled the central region the``hot corona'' and will refer to as such in the present paper too.   The temperature of the hot gas within the corona is a factor of $\sim 2-2.5$ greater than the system's virial temperature at $z=0$ \citep{BBP:1999,Voort:2011} of 
\begin{equation}
T_{vir}\approx 4\times 10^5\; {\rm K} \ \left[\frac{M_{halo}(z)}{7.6\times 10^{11}\; M_\odot}\right]^{2/3} (1+z),
\end{equation}
where we have scaled the relationship to \eris's dark matter halo mass at $z=0$.
The corona has a radius of about 100~kpc at $z=0.5$ and grows to 140~kpc at $z=0$, equivalent to 0.6~$r_{vir}$. Its mass is approximately $7\times 10^9\; M_\odot$ while that of the warm-hot gas is about four times larger, i.e.~$\sim 3\times 10^{10}\; M_\odot$. The combined properties of the hot corona and the warm-hot atmosphere in \eris~are in excellent agreement with a number of different observational constraints and characteristics of the Milky Way halo gas, including its electron density profile and the X--ray luminosity of $\sim 10^{39}$~erg/s
in the 0.5-2~keV band \citep[see][ and references therein for details]{Sokolowska:2016}.   

Here, we start by assessing \venus. {As we did with \eris,} we compute the X-ray luminosity of the gas particles  within the virial radius of a halo using the radiative rates from the Astrophysical Plasma Emission Code3 (APEC) \citep{APEC}, which assumes an optically thin gas in collisional ionization equilibrium (for more details on the calculation, see Paper~I).  {We find that the X-ray luminosity of \venus, $L_X=1.2\times10^{39}$~erg/s, is also in excellent agreement with the Milky Way observations. }

Figure~\ref{fig:temp_map} compares the \eris~and \venus~gas temperature maps at three different time steps: $z=4,\; 3,\; {\rm and}\; 0.5$.  {As mentioned previously, the configuration of the filamentary network in the two simulations is different, resulting in two distinct evolutionary sequences for the gas. For example, }at $z=4$, cold filamentary flows penetrate the halos of the  of the main progenitors in both runs {but gas in \eris~is hotter.  At $z=3$, a central region containing both hot and warm-hot gas 
has formed about the most massive progenitor galaxy in \eris~while in \venus, the gas has yet to heat up to the same temperatures and is also more widely distributed among the multiple progenitors.  By $z=0.5$}, however, \venus~ acquires a corona/gaseous halo like that in \eris.  

That both \eris~and \venus~reach similar endpoints with respect to the structure of gaseous halo/corona, even though they do so via 
different evolutionary pathways, raises a number of questions regarding the processes that govern their evolution and facilitate the convergence at $z=0$.
%{The existence of similar diffuse halos that appear in different environments opens a question of what governs their evolution, and how they arrive at this end-state.} 
In what follows, we analyze the simulations to identify the key processes involved.  The one phenomenon that stands out is powerful episodic galactic outflows.  We also discuss the impact of the outflows and the growing hot coronae on other structures in the galactic halo, including the cold filaments depositing gas onto the central galaxy, and 
%investigate which processes are essential for explaining their origin.  {Then, we focus our attention on the role of supernova (SN) feedback in shaping these halos.} Finally, we discuss the impact of growing hot coronae 
on the transition from cold filamentary to hot quasi-spherical accretion mode.

Throughout this paper, we follow the schema adopted in Paper I and 
discuss the distribution and the properties of gas in different temperature ranges, namely
cold gas ($T<3\times10^4$~K), warm gas ($3\times10^4{\rm K}<T<10^5{\rm K}$), warm-hot gas ($T=10^{5-6}{\rm K}$), and hot gas ($T>10^6\rm K$).  
We do so to facilitate comparison with the observational literature where the discussion is informed by the diagnostics used to study the gaseous halos \citep[see the review by][ for details]{Putman:2012aa}.  For completeness, we note that the first two categories correspond to gas probed via  HI and H$\alpha$ emission lines, as well as by various low and intermediate ion absorption lines, including Ly$\alpha$, Mg II, S III, S IIII, C II, CIII, and O I.  The third category corresponds to gas studied via C IV and O VI absorption lines, and the fourth to X--ray emissions and O VII and O VIII lines.

\subsection{Growth of the diffuse medium: an overview} \label{sec:overview}

\begin{figure*}
\hspace{-.5cm}
\includegraphics[scale=0.28]{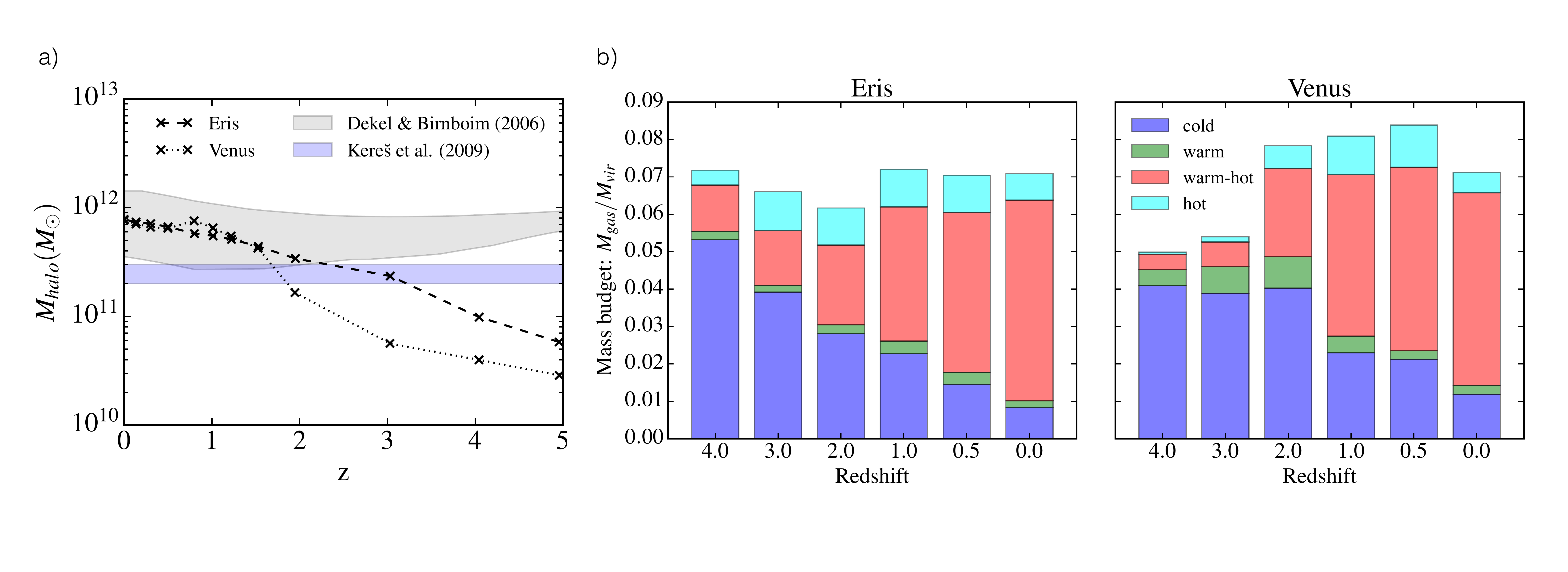}
\caption{ {The stable shock condition and the formation of the gaseous halos in \eris~and\venus.} a)~Mass of the most massive progenitor halo as a function of redshift is compared with the critical mass for the development of a stable shock. The grey band represents the metallicity-dependent expectation for the critical mass necessary to develop a stable inner shock at 0.1~$r_{vir}$\citep[][; metallicity range 0-0.3~$Z_{\odot}$]{Dekel:2006}.  {The blue band marks a mass range of halos from \citet{Keres:2009} that can sustain atmospheres of hot, virialized gas.} b)~Gas mass budget of the gaseous halos 
around the galaxies within the most massive progenitor halos in \eris~and \venus~ at various redshifts, measured within their virial radii. Cold gas: $T<3\times10^4$~K; warm gas: $3\times10^4{\rm K}<T<10^5{\rm K}$; warm-hot gas: $T=10^{5-6}{\rm K}$; hot gas: $T>10^6\rm K$.}
\label{fig:form_vs_mcrit}
\end{figure*}

\arif{The conventional description \citep[see, for example, ][]{Keres:2009}
of the formation of the diffuse gaseous halos and their X-ray luminous coronae assumes that they are primarily generated by hot mode accretion {\it onto the halos} once the halos' mass exceeds the critical mass threshold;
that is, the accreting gas is shock-heated by a stable accretion shock at the virial radius. In the post-shock region, the gas that enters at an earlier time is pushed  deeper into the halo by presently inflowing gas.   If the cooling time for this diffuse gas is longer than the compression time, the gas heats up as it is squeezed deeper into the halo.  \citet{Joung:2012} find that in their simulated MW-like galaxies, the dominant component of this inward flowing diffuse gas is $< 10^5$K beyond the virial radius, $10^5 < T < 10^6$ K (i.e.~warm-hot) down to $\approx 50$ kpc, and then $>10^6$ K (i.e.~hot) within the innermost $20 - 50$ kpc.  }
Here, we assess this picture in the context of our simulations.}
%test the validity of this picture in our simulations. First, we investigate when halos of our galaxies enter the critical mass regime; then, we compare the resulting redshift with the time, during which hot diffuse atmospheres of gas begin to appear around those halos.}

In the left panel (a) of Figure~\ref{fig:form_vs_mcrit}, we show the evolving total (virial) halo mass for the most massive progenitor halo  {in \eris~ and \venus~} as a function of redshift. The grey band shows the \citet{Dekel:2006} metallicity-dependent critical mass threshold that they argue a halo must grown to in order to sustain an inner shock at 0.1~$\rm r_{vir}$.  The band's width is due to its dependence on metallicity and corresponds to a spread of 0.03$\rm Z_{\odot}$ to 0.3$\rm Z_{\odot}$.  In this model, the critical mass for sustaining a stable virial shock {(at $\rm r_{vir}$)} exceeds $10^{12}M_{\odot}$,  even with $\rm Z=0.03Z_{\odot}$, and hence sits above the grey band.  
\arif{The blue band corresponds to the nearly redshift independent (at least for 
$0 \leq z \leq 3$) threshold mass range of $\rm 2-3 \times 10^{11}~M_{\odot}$, where halos start to develop diffuse gaseous atmospheres in the simulations of  \citet{Keres:2009}. This is also mass range where the accretion onto the halos transitions from predominantly cold to predominantly hot mode.  
%We note, however, that \citet{Keres:2009} did not include metal-line cooling in their simulations.  \arifhum{NEED TO REMOVE THIS -- COMPLICATES DISCUSSION FURTHER IN: And although not shown in the figure, we note that \citet{Voort:2011}, who allow for both self-consistent enrichment of the gas   and metal cooling in their simulations, find a slightly higher and mildly redshift-dependent threshold mass, decreasing from $1.6\times 10^12\;M_\odot$ at $z\approx 3$ to  $1\times 10^12\;M_\odot$ at $z\approx 2$, and finally to $7\times 10^11\;M_\odot$ at $z\approx 0$.}  
Given their different mass accretion histories, it is not surprising that  \eris~and \venus~simulations transition to the hot mode accretion-dominated phase at different times.   \eris~crosses into the \citet{Keres:2009} blue band at $z=3$ and into the grey band at $z=2$.  
%\arifhum{AND THIS: and passes the \citet{Voort:2011} threshold at $z\approx 0.3$.} 
\venus~ lags \eris, crossing the blue and grey bands at $z=2$ and $z=1.5$, respectively.  
%\arifhum{AND THIS: It, however, reaches the \citet{Voort:2011} transition mass at $z\approx 1$. }

  In the middle and the right panels (b) of Figure~\ref{fig:form_vs_mcrit}, we show the gas mass budget within the virial radius of the most massive progenitor halos of \eris~ and \venus, respectively, at 6 different redshifts.  The gas mass is normalized to the virial mass of the halo at the redshifts under consideration and categorized by temperature: cold ($T<3\times10^4$~K), warm ($3\times10^4{\rm K}<T<10^5{\rm K}$), warm-hot gas ($T=10^{5-6}{\rm K}$) and hot ($T>10^6\rm K$).   We explicitly exclude gas in the galactic disk, i.e. gas with density higher than the star formation density threshold (5 atoms/cc), in order to isolate the gaseous halo.  The differences in the gas fractions between the two simulations is due to their different assembly history; i.e., the relatively low gas fraction before $z=2$ in \venus~ is correlated with its late halo assembly.  However, after \venus~undergoes its last major merger and the halos masses in the two simulations converge, the gas fractions in both runs become similar too.
}
  
\arif{
  Examining the distribution of gas across the different phases in \eris~first, we find that although the cold gas (blue) is the most abundant component until $z=2$, there is already significant warm-hot, and especially hot, halo gas by $z=3$.  The two components combined comprise $\simeq 30\%$ of the total diffuse gas budget at $z=4$ and $\simeq 40\%$ at $z=3$.
  %, which is somewhat earlier than in  \citet{Keres:2009} even though the cooling in our simulations is more efficient, 
  In \venus~the hot and warm-hot diffuse gas comprises  
$\simeq 20\%$ at $z=3$ and $\simeq 40\%$ by $z=2$.  
%These results are similar to those of \citet{Keres:2009} especially once the slight differences in the cooling physics between the two sets of runs is accounted for.
Of particular interest to us is the presence of nearly constant fraction of hot ($>10^6$) gas at $0.5 \lesssim z \lesssim 3$ in \eris~ and $0.5 \lesssim z < 2$ in \venus. This fraction is $14-16\%$ in \eris~ and $12-13\%$ in \venus. The temperature of this gas is a factor of $\sim 2-3$ higher than the virial temperature of the halos at these epochs: $T_{vir}=(5.3, 7.5, 7.1, 6.8, 6.1) \times 10^{5}$~K for \eris~ and $(2.9, 2.8, 4.5, 7.4, 6.0) \times 10^5$~K for \venus~ at $z=(4,3,2,1,0.5)$, respectively.
}
Although this hot gas is not the most abundant gas phase, it is primarily responsible for the extended diffuse X-ray emissions and is also the  
component that is traced by the  OVII/OVIII absorption/emission.  As we summarized in the Introduction and discussed in much greater detail in Paper I, the low-z hot gas properties in our simulations agree remarkably well with the observations. 

%\arifhum{
%ARIF COMMENT:   I HAVE NO IDEA WHAT ALEX WAS WRITING HERE.   IF I COMPARE TO KERES ET AL, ERIS CROSSES THE 50-50 THRESHOLD AT z=3 AND THIS IS WHERE WE SEE 40\% WARM HOT+HOT.   THIS IS THE BAND WHERE KERES ET AL SEE 50\% GAS WITH T > 2.5E+5 K SO THERE TEMP CUT IS HIGHER THAN OUR (1.E+5).   WE ARE AHEAD OF VOORT ET AL.  BUT THEY USE METAL LINE COOLING AND I AM STARTING TO REALIZE (AND NOW, CONFIRMED) THAT ERIS DOES NOT, EXCEPT FOR THE COLDEST GAS.  SO AS FAR AS OVERALL MASS SCALES GOES, WE AREN'T THAT DIFFERENT FROM KERES ET AL.  I HAVE MODIFIED THE ABOVE TO PUT THE EMPHASIS SQUARELY ON THE HOT GAS, WHOSE TEMP EXCEEDS THE VIRIAL TEMP.}

%This implies that a significant fraction of diffuse gas (red and cyan bars in the panel (b) of Figure~\ref{fig:form_vs_mcrit}) should appear close to these redshifts. However, this is not what we observe.
%Although the cold gas is the most abundant gas phase until $z=2$, both halos can sustain hot and warm-hot atmospheres earlier than that. Taking the example of \eris, already at $z=4$, the diffuse gas (warm-hot and hot combined) amounts to 40\% of cold gas; at $z=3$, to 70\%. Diffuse gas appears in \venus way before the halo reaches the critical mass as well -- at $z=4$, it amounts to 12\%, at $z=3$ to 25\% and at $z=2$ to 75\% of the cold gas, respectively. These results imply that hot accretion cannot be the source of diffuse halos developing at early redshifts.

\begin{figure}
\centering
\includegraphics[scale=0.56]{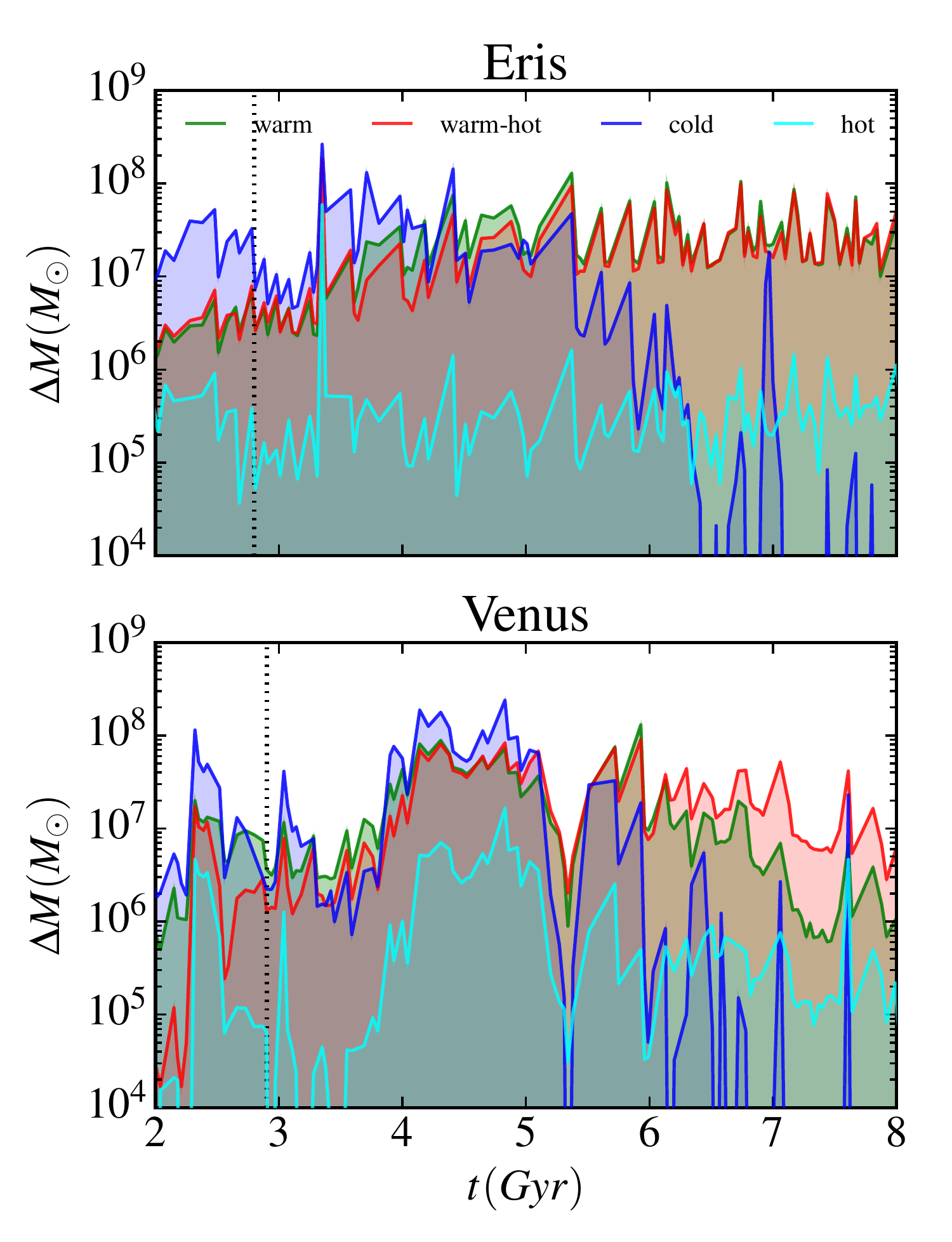}
\caption{ {Temperature of gas particles present in a $z=0$ hot corona at first $\rm r_{vir}$ crossing, }showing two-stage halo gas evolution. The dotted line marks approximately the time of the equivalence of the cold and hot mode accretion ($z=2.5$).}
\label{fig:coronalhistory}
\end{figure}

To identify the origin of the hot gas, we flag the gas particles within the virial radius that comprise this component at $z=0$, and track their temperatures back in time. We also note the time when they accreted onto the most massive progenitor halo in the two simulations.
%Our diffuse halos match well these constraints, therefore we investigate under what conditions this agreement is achievable. We select particles, which are hot ($T>10^6$~K) at $z=0$ and at distances smaller than the virial radius, and trace their temperature and their time of accretion onto a halo. 
We treat a particle as accreted when it crosses the virial radius of the halo for the first time.  

In Figure~\ref{fig:coronalhistory}, we show that the gas that  ends up in a hot coronae at $z=0$ is not accreted exclusively at late epochs. A significant fraction (i.e. $\sim 30$\%) of it is accreted at early times, over the first $5-6$ Gyrs, and is cold at the time.  This cold-accreted gas, which comes in via the filamentary streams and is, early on, primarily deposited onto the central galaxy must be heated to over a million degrees and expelled from the galaxy in order to become a part of a hot corona.  There are only two main mechanisms that can do this:  strong shock-heating and dispersal of the gas as a result of major mergers \citep{Poole:2006,Poole:2008,Sinha:2009}
and/or heating and expulsion via supernova-driven galactic outflows
\citep[e.g.~][]{
Oppenheimer:2006,Dave:2011,Dave:2012,Voort:2011,Voort:2016,
In-N-OutBaryons:2016,Stewart:2017}.  We note that we write ``and/or'' because mergers, especially gas-rich mergers, are frequently associated with starbursts.  

The contribution of the cold-accreted gas to the coronae drops off steeply after $\approx 5.5$~Gyrs of evolution (i.e. after $z\sim 1.15$) in favor of warm and warm-hot accretion in both simulations. By this time, both halos are of similar mass and growing at nearly the same rate.  The warm/warm-hot accretion flow never settles onto the central galaxy and the fraction that contributes to the corona is gas that is initially shock-heated upon accretion and then heated to coronal temperatures either via compression or via additional shocks in the inner halo due to interactions with the outflowing winds.    

Figure~\ref{fig:coronalhistory} also shows the time when the amount of cold-accreted gas \emph{first} equals the amount of gas coming in via warm-hot and hot modes.  This is delineated by the black dotted line (and referred to as the epoch of ``equivalence'') and corresponds to approximately $z=2.5$ for \eris~ and $z=2.3$ for \venus.  We did not expect that these times would be comparable given that the mass of the dominant halos in the two runs is different by a factor of a few at the time (see Figure~\ref{fig:form_vs_mcrit}a).  However, from about $z=2.9$ to $1.4$, \venus~ experiences a series of gas-rich mergers.   Each merger leads to a sharp increase in the diffuse gas fraction, resulting in an overall accelerated growth to similar level as in \eris~ by $z=2$ (panel b of Figure~\ref{fig:form_vs_mcrit}).  

%Hence, cold-accreted gas must be either shock-heated via mergers and end up in a hot state with a very long cooling time, or it must be kept hot by the energy of supernovae (or both). It could also become a part of outflows and return as second-generation hot accretion at later t. The exact cause of the generation of hot coronae is the main topic of the paper, and we focus on this problem in the next section.}

%\arifhum{I AM SPLITTING THIS SECTION HERE AND MAKING A SEPARATE SUBSECTION.  SEEMS APPROPRIATE.  LET ME KNOW.}\\

\subsection{The distribution and thermal properties of the gaseous halo}\label{sec:tform}

 {In this subsection, we investigate the distribution of diffuse gas that comprises the gaseous halo, and characterize its thermal evolution as a function of time.  As previously, we ignore the gas that belong to the disk, i.e. all the gas with density higher than the star formation density threshold of $5$ atoms/cc. }

\begin{figure}
\centering
\hspace{-.7cm}
\includegraphics[scale=0.6]{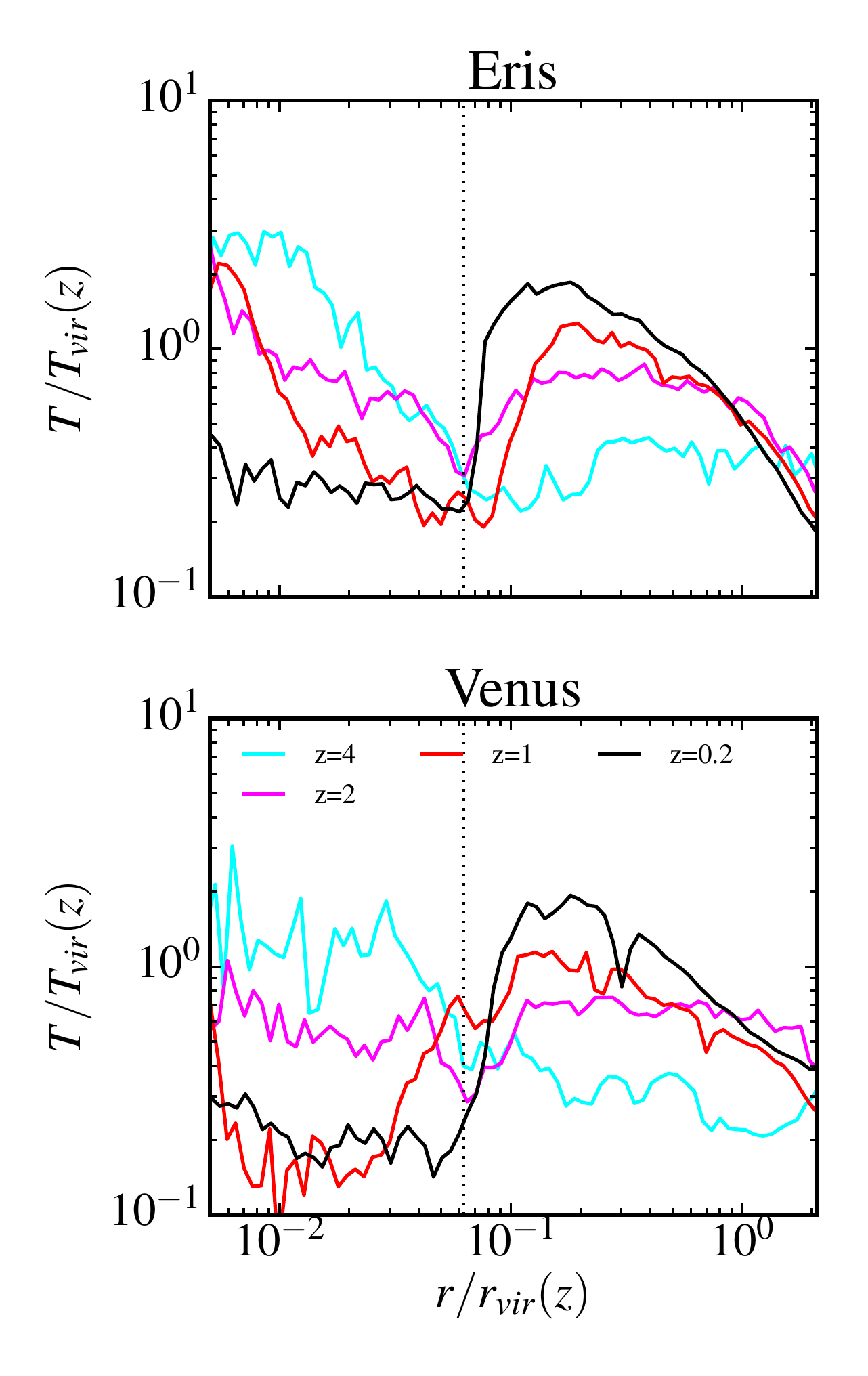}
\caption{Radial distributions of gas temperature normalized to the virial temperature of the halo at a given redshift, showing the diffuse halo in formation. The vertical line marks the approximate extent of the disk at $z=0$. Top to bottom: \eris, \venus.}
\label{fig:temp}
\end{figure}

 {Firstly, we present the temperature profiles of the gaseous halo in Figure~\ref{fig:temp}. This is mass-weighted average temperature of the gas in radial shells around the galactic center and normalized to the virial temperature of the halo: $T_{vir}=(5.3, 7.5, 7.1, 6.8, 6.1) \times 10^{5}$~K for \eris~and $(2.9, 2.8, 4.5, 7.4, 6.0) \times 10^5$~K for \venus~ at $z=(4,3,2,1,0.5)$, respectively.  One can distinguish two regions in these profiles, to the right and left of the black vertical line ($\rm r/r_{vir}=0.06$), which marks the approximate extent of the galactic disk at $z=0$. The gas to the left cools to a fraction of the virial temperature, flows into and supplies fresh gas to the galactic disk.  The gas to the right grows hotter with time.  Prior to $z\approx 2$, the gas is nearly isothermal across this region but thereafter, the temperature distribution has a maximum near $0.2r_{vir}$ ($40$~kpc at $z\approx 0$) and declines to $\simeq 0.5 T_{vir}(z)$ at the virial radius. \arifho{Such an outwardly declining temperature profile is normal feature of the hierarchical structure formation models} \citep[see, for example,][]{Lewis:2000}. At late times, the peak temperature increases to approximately twice the virial temperature.  The particular shape of $ T/T_{vir}$ suggests that some heating process (or, a combination therefore) localized to the inner halo or related to the galactic disk is responsible for the hot corona, which, as we have mentioned previously, cocoons the galaxy and extends out to  $0.6~r_{vir}$ at low redshifts.  
}

\begin{figure}
\centering
\hspace{-.7cm}
\includegraphics[scale=0.6]{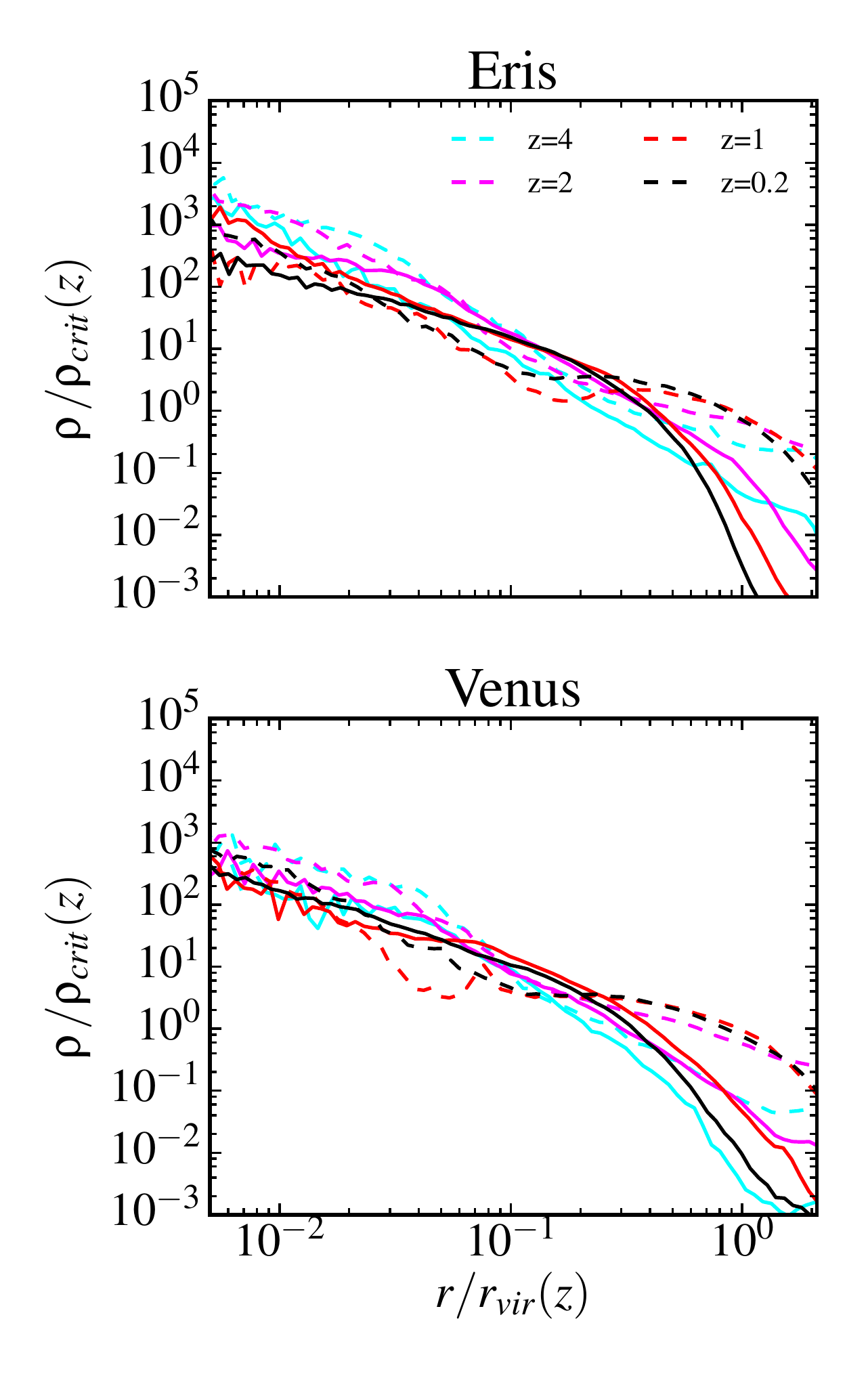}
\caption{Radial density distribution of gas and its evolution with redshift. Dashed and solid lines denote the warm-hot and hot components of gas, respectively.  Quantities are normalized to the critical density and the virial radius of the galaxy at a given redshift, and show the development of the onion-like structure: a corona embedded in the warm-hot soup of gas.  }
\label{fig:dens}
\end{figure}

 {Next, we look specifically at the spatial distributions of the  hot ($>10^6$~K) and warm-hot ($10^{5-6}$~K) components of the gaseous halos.
In Figure~\ref{fig:dens}, we compare their density profiles at different redshifts.  At first ($z=4$), warm-hot and hot gas profiles in both simulations fall off as a power-law with increasing radii and the density of the warm-hot gas is higher. 
After $z=2$, however, several differences arise: First, although the power-law index of the warm-hot gas profile at small radii remains approximately unchanged, its amplitude decreases.  It also develops a shallow trough with a minimum at $(0.1-0.2)\; r_{vir}$ and then
%Around $0.1 r_{vir}$, the profile flattens to form a trough at about $  0.2 r_{vir}$, then 
rises slightly to a local maximum before dropping off at large radii.  At the same time, hot gas profile flattens slightly to $ 0.4\; r_{vir}$ before dropping off steeply.  This steep drop-off indicates that  hot gas can be found only within a fraction of a virial radius ($\sim 0.6-0.8\; r_{vir}$) and not beyond. 
Additionally, the evolution of the two profiles jointly results in a pocket between $\sim 0.05\; r_{vir}$ and $\sim 0.5\; r_{vir}$ (\eris) and 
$\sim 0.02\; r_{vir}$ and $\sim 0.2\; r_{vir}$ (\venus)
where the hot gas concentration exceeds that of the warm-hot gas.  This pocket emerges at $z\simeq 2$ in \eris.}

\begin{figure}
\centering
\hspace{-0.72cm}
\includegraphics[scale=0.6]{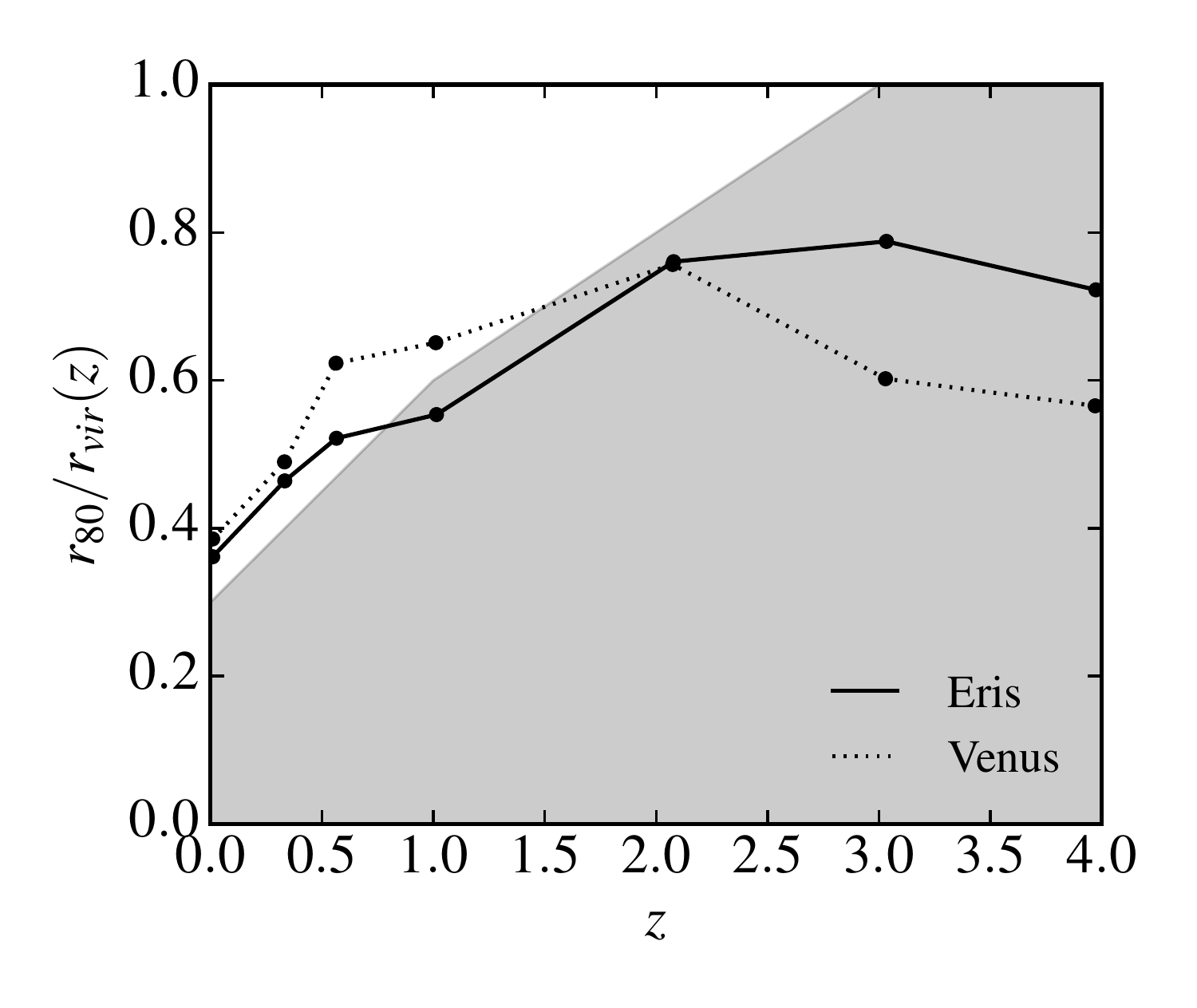}
\caption{The size evolution of hot gas halo. $r_{80}$, defined as a radius encompassing 80\% of its mass within the virial radius, normalized to the virial radius ($r_{vir}$). Grey-shaded region indicates where cooling of hot gas is still important (i.e. $t_{cool}<<t_{Hubble}$).}
\label{fig:r80}
\end{figure}

We quantify the extent of the hot gas as a function of time in Figure~\ref{fig:r80}.  We define its extent as a radius encompassing 80\% of the total hot gas mass within the virial radius ($r_{80}$) and find that before the formation time of the corona (\i.e.~around $z=4-3$), the hot gas extends out to 80\% of the virial radius. As the halo accretion mode switches to hot mode, the hot gas component in the halos is increasingly confined by the accumulating blanket of warm-hot gas, to $0.6\; r_{vir}$ at $z=2$ and decreasing with time to $0.4\;r_{vir}$ by $z=0$.

\arif{
The grey-shaded area in Figure~\ref{fig:r80} shows the radius within which the cooling time of the hot gas is shorter than the Hubble time, as a function of the redshift.  To estimate this cooling radius, we calculated cooling time as a function of radius according to $t_{cool} = U/\Lambda$, where both the internal energy, $U$, and the radiative cooling rate, $\Lambda$, are taken from the simulations.
We have verified that in both simulations, the cooling time is very short near the galactic center (e.g., within 0.1~$r_{vir}$ it drops to orders of magnitude below 1~Gyr even at $z=1$), and monotonically increases with radius.  
Nonetheless, the cooling time of all of the hot gas within the virial radius at $z=3$ is less than 1~Gyr time and that of all the hot gas within, for example,  $0.8\; r_{vir}$ at $z=2$, $0.6\; r_{vir}$ at $z=1$, and $0.3\; r_{vir}$ at $z=0$ is less than the Hubble time.  This means that there shouldn't be any hot gas in the central regions unless it was either being continually replenished or reheated.   Shock-heating of the gas as a result of major mergers or heating/replenishment by galactic outflows would both do the trick. 
}

\subsection{Role of feedback and mergers} \label{sec:feedback}

\begin{figure}
\centering
\hspace{-0.6cm}
\includegraphics[scale=0.55]{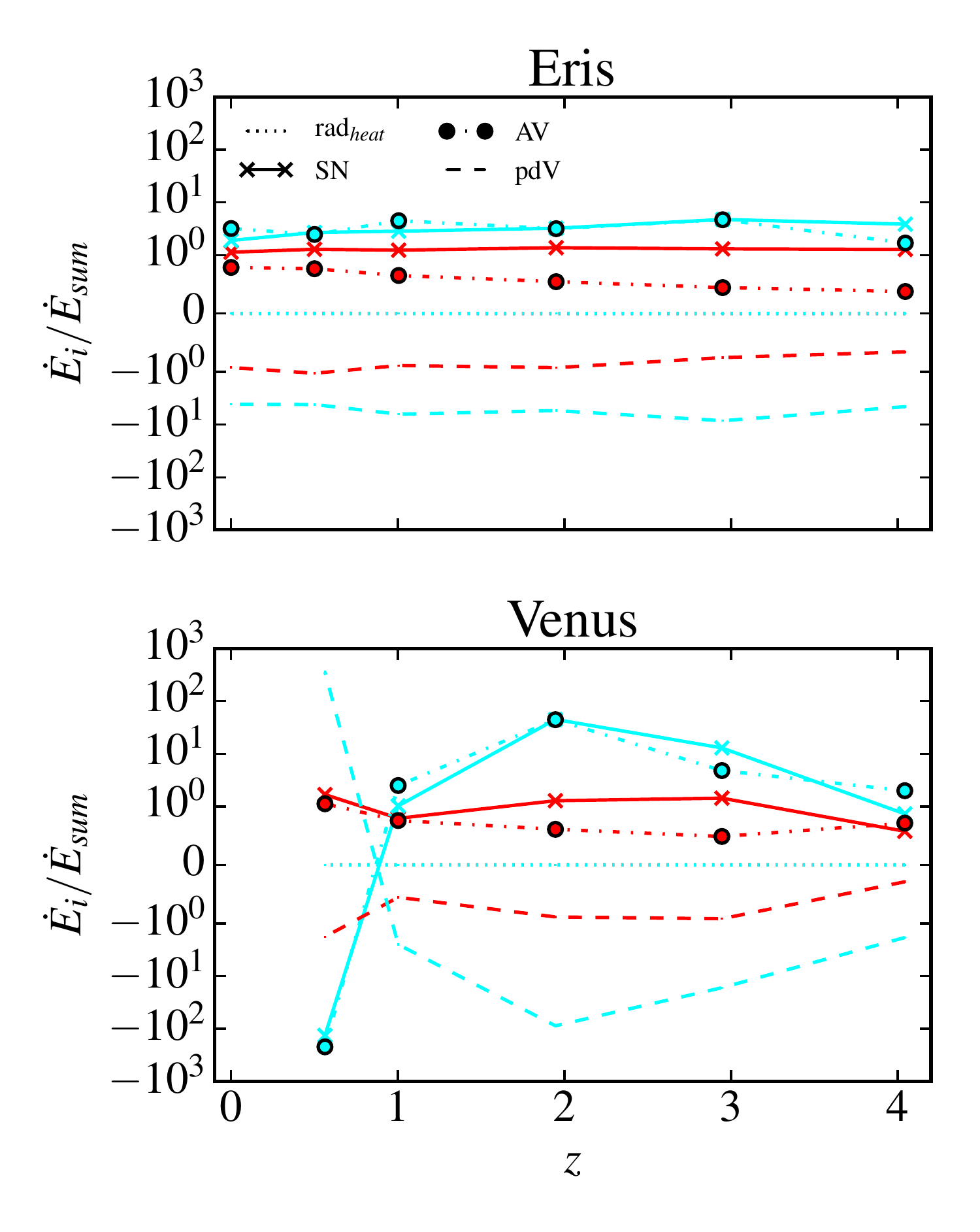}
\caption{Energy injection rates for various gas phases in the region encompassed by a sphere of a virial radius around a galaxy, normalized to the total heating rate. The color coding corresponds to the warm-hot (red) and hot (cyan) gas phases. Legend: $rad_{heat}$ -- heating due to atomic/radiative processes only, $SN$ -- supernovae thermal feedback, $AV$ -- artificial viscosity, $pdV$ -- work done by/on the gas.}
\label{fig:energy}
\end{figure}

{In this section, we attempt to identify the major heating mechanisms that contributes to the formation and build-up of the hot corona.
The design of our simulation code, {\sc Gasoline}, allows us to determine the impact of each of the following heating mechanisms on the diffuse halo gas:
radiative heating ($\dot{E}_{rad}$), supernovae thermal feedback ($\dot{E}_{SN}$), and artificial viscosity ($\dot{E}_{AV}$), which reflects heating due to  shocks, mixing etc.  Here, ``shocks'' refers not only to accretion shocks but to shocks in general, including those triggered by mergers, galactic outflows, etc. We also show the heating (cooling) due to adiabatic compression (expansion) ($\dot{E}_{pdV}$).  While the first three $\dot{E}_{i}$ are non-negative, the latter can be positive (indicating heating due to compression) or negative (indicating cooling due to expansion).   

We focus on two particular phases of gas: the warm-hot gas ($T=10^{5-6}{\rm K}$) and the hot gas ($T>10^6\rm K$). We compute the mass-weighted $\dot{E}_{i}$ for the different heating mechanisms for the gas within the virial radius as a function of redshift, and normalize all energy injection rates to the total heating rate $\dot{E}_{sum}=\sum_i \dot{E}_i$, where $i$ stands for each of the processes mentioned above. 
}

\arif{
The red and cyan lines in Figure~\ref{fig:energy} show the results for the warm-hot and hot gas, respectively.   To start with, we note that the impact of the radiative heating on the two phases is negligible.   We also find that both phases are expanding (and hence, cooling) at all redshifts.
This is true even at the lowest redshift in \venus~despite the dashed cyan line's sudden steep rise to the positive domain.   As we have mentioned previously, \venus~undergoes a major merger at $z=0.9$ and in the aftermath, the hot gas expands and cools so rapidly that even the total energy injection rate $\rm \dot{E}_{sum}$ becomes negative and consequently, $ \dot{E}_{SN}/\dot{E}_{sum}$, $ \dot{E}_{pdV}/\dot{E}_{sum}$ and $ \dot{E}_{AV}/\dot{E}_{sum}$ all change signs. As for the heating mechanisms, Figure~\ref{fig:energy} shows that both SNe heating and shocks/mixing contribute equally to the heating of the hot phase across all redshifts while SNe heating is slightly more important for the warm-hot gas.   We reiterate that shock heating is not exclusively due to shocks associated gravitational processes (i.e. accretion and mergers) but also includes contributions from shocks engendered by galactic outflows.   In other words, $\dot{E}_{SN}$ represents the minimum amount of energy injected by the SNe.
\emph{Taken together, the above findings show that SNe heating and subsequent inside-out expansion play a pivotal role in the the build-up of the hot corona and the warm-hot phase in our MW-like systems.}
}

%The heating of the warm-hot gas is dominated by supernovae across all redshifts. The second-most important heating source of that phase is artificial viscosity, with a rate that is lower within a factor of two.} The build-up of the hot phase, however, is aproximately equally driven by both of these heating agents, as is shown with the cyan lines. We note that, as AV is only a tracer of shocks and not the measure of shock heating in these simulations, we treat it as such. The core result of this analysis is thus that, since supernovae heating is more, or at least equally, important source of heating, it must play a major role in shaping the diffuse halos of our galaxies.  

\begin{figure*}
\centering
\hspace{-0.05cm}
\includegraphics[scale=0.5]{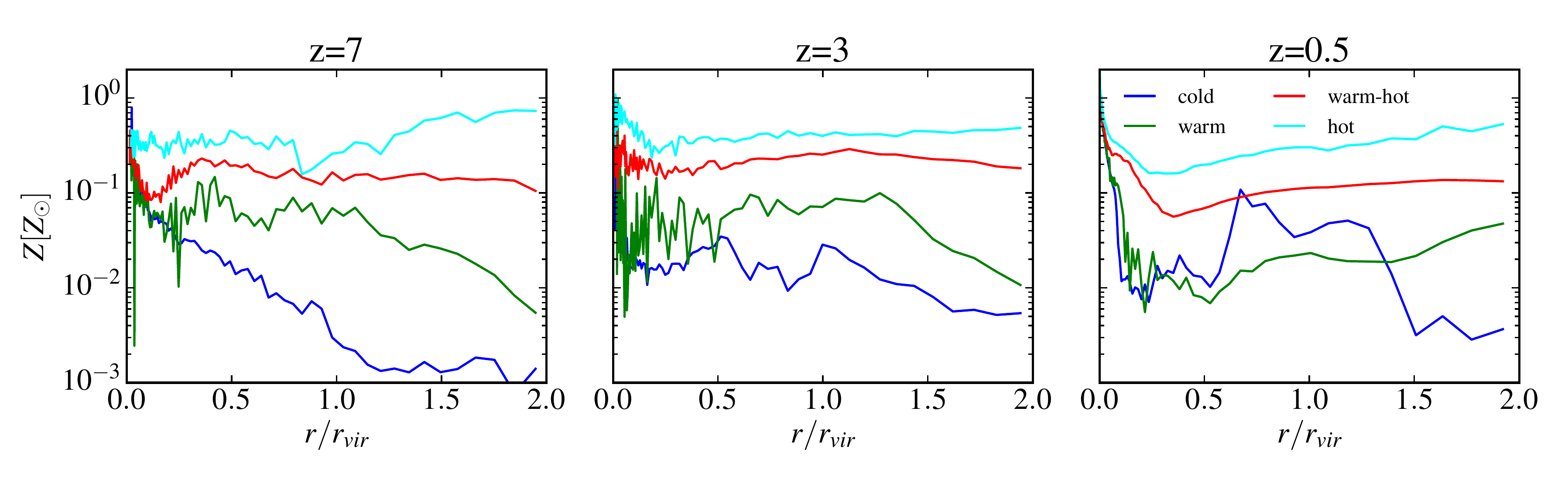}
\caption{Example of radial metallicity profiles per gas phase at 3 different time steps. Case study: \eris. Profiles of \venus are very similar and show the same trend: hot gas is the most metal-enriched gas phase, and warm-hot gas is the second-most metal-enriched gas phase.}
\label{fig:metal}
\end{figure*}

{Yet another feature that points to the importance of feedback and galactic outflows is the high metallicity of the warm-hot and the hot diffuse gas from as early as $z=7$.   
%both as early as $z=7$ and late as $z=0.5$. 
Figure~\ref{fig:metal} shows the radial metallicity profiles (in solar units 
assuming $Z_{\odot}=0.0194$ \citep{Anders:1989}) for different phases of gas at three epochs, $z=7$, $z=3$, $z=0.5$, in the \eris~simulation.  The trends in \venus~are very similar.    As in the previous section, we are interested in the metallicity of the diffuse gas only and not in the disk; therefore, we only consider gas below the star formation threshold density. The results show that the hot and the warm-hot phases are the most and the second-most metal rich phases, respectively, with a nearly constant metallicity, within the range $0.3-0.5\; Z_\odot$for the hot phase and $0.1-0.2\; Z_\odot$ for the warm-hot phase, at least as far out as $2r_{vir}$ and as early as $z=7$.  The metals in our simulations come from SNI and SNII within galaxies and galactic outflows is the only available mechanism for transporting these out into the halo and beyond \citep[c.f.~][]{Dave:2012,Rahmati:2016MNRAS}.   The difference in the metallicity of hot and the warm-hot gas is the result of the latter is a mixture of high-metallicity galactic outflows and cosmological influx of low metallicity gas that accretes onto the halo.  We also note that there is a inward rise in the profiles developing close to the halo center at $z=3$, which suggests that the metal-rich outflows are increasingly confined at late times as the potential well deepens and the gaseous halo itself grows.   
}

\begin{figure*}
\centering
\includegraphics[scale=0.46]{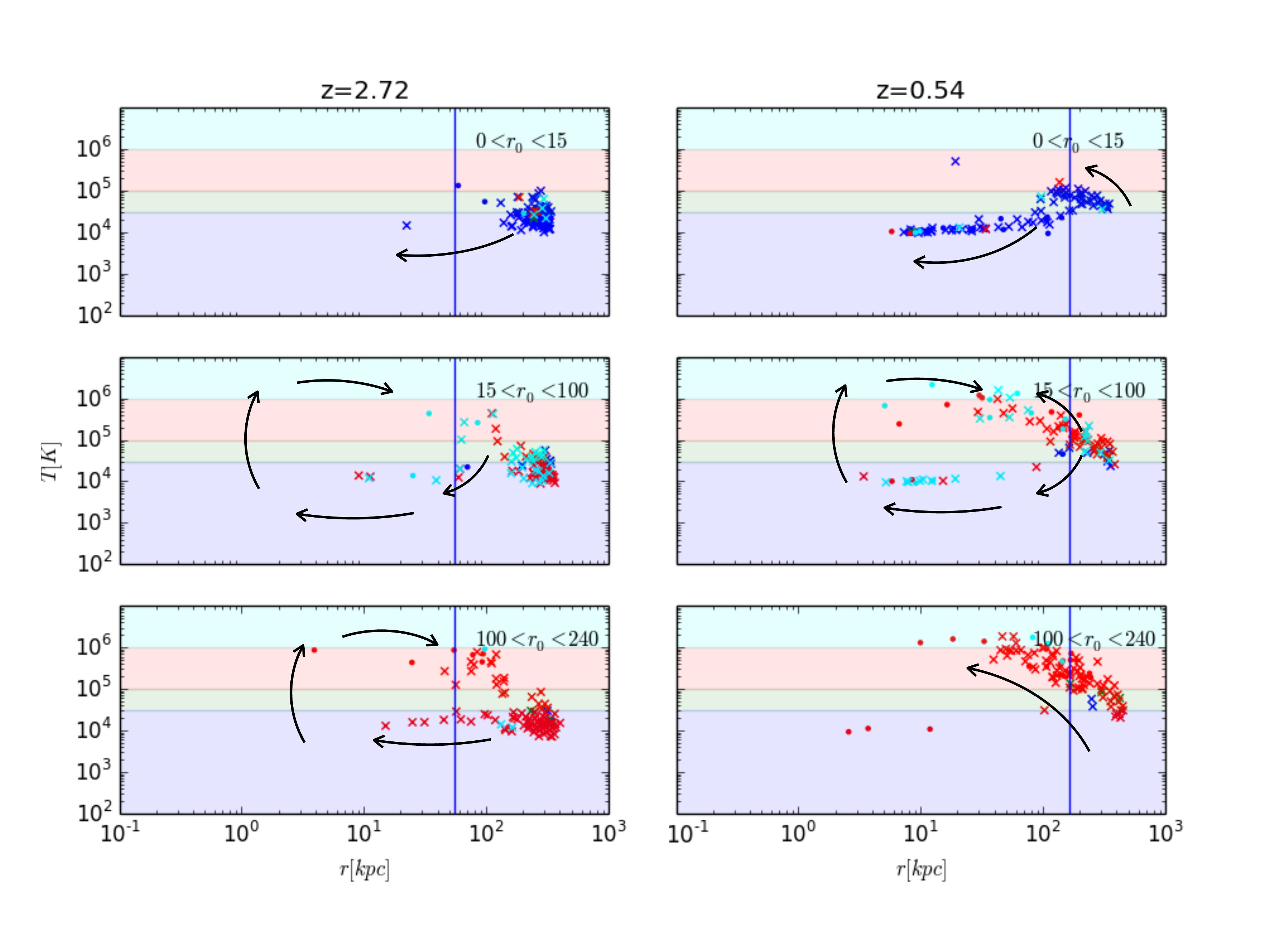}
\caption{The time evolution of a 100 particles, in the radius-temperature space, from each of the three radial zones of the $z=0$ gaseous halo in \eris~(c.f.~Table \ref{tab:massfrac}).  The left and right columns show the state of the particles at $z=2.72$ and $z=0.54$, respectively. 
The top, middle and the bottom rows show the results for the innermost  ($r_{0}<15$~pkpc), the middle ($r_{0}\in(15,100)$~pkpc), and the outer zone 
($r_{0}\in(100,240)$~kpc).   The background of each panel is color-coded according to the temperature range of the different phases of gas under consideration (hot, warm-hot, warm, cold) and color of the individual points similarly indicates the phrase they end up in at $z=0$.  The arrows mark the typical trajectories of particles at the epochs shown.  The individual particles are represented with a dot or a cross, depending on whether they have receive feedback energy at any time prior to the redshift shown, or not (cross = no). The vertical line marks the virial radius. For a detailed description of the experiment, including the selection method, see the text. Case study: \eris.}
\label{fig:onion}
\end{figure*}

\arif{To further understanding the origin of the warm-hot and hot gas in the diffuse halo, we divide the gaseous halo in \eris~into three zones: (i)~disk-dominated ($r_{0}<15$~kpc), (ii)~corona-dominated ($r_{0}\in(15,100)$~kpc), and (iii)~the outer warm-hot reservoir ($r_{0}\in(100,240)$~kpc).  We determine the fraction of gas mass in the different phrases in each zone.   The results are shown in Table \ref{tab:massfrac}.  We then select 100 particles from each zone, distributed across the different phases in accordance with the mass fractions (e.g., the representative set of particles for the central region comprises 90 cold, 1 warm, 4 warm-hot and 5 hot particles), color-code the particles by their $z=0$ phase, track them back in time, and follow their evolution forward in time on the radius-temperature ($r-T$) diagram.  Two such diagrams, for $z=2.72\; {\rm (left\; column)\; and}\; 0.54 {\rm (right\; column)}$, are shown Figure~\ref{fig:onion}, where we indicate the typical trajectories of particles with black arrows; an animation (movie) illustrating the time-evolution in a more visually informative format is included with the supplementary material.

\begin{table}
\vspace{5.0pt}
\centering
\begin{tabular}{l|c|c|c|c}
Zone & Cold & Warm & Warm-hot & Hot \\
\hline
$r_{0}<15$~kpc			& 90\%	& 1\%	& 4\%	& 5\%	\\ 
$r_{0}\in(15,100)$~kpc	& 9\%		& 0\%	& 58\%	& 33\%	\\ 
$r_{0}\in(100,240)$~kpc & 2\%		& 3\%	& 92\%	& 3\%	\\
\end{tabular}
\vspace{10.0pt}
\caption{The mass fraction of gas in the four phases in each of the three zones, at $z=0$.}
\label{tab:massfrac}
\end{table}
}

\arif{
This experiment allowed us to distinguish three different patterns of accretion and their evolution.  These can be categorized into two types, with the left column in Figure~\ref{fig:onion} being an example of the first and the right column of the second type: 

We first consider the innermost zone.   Not surprisingly, the gas enters the halo via cold flows at early times and flows onto the central galaxy. At late times, the gas enters the halo and is shock-heated to $T\sim 10^5$~K close to the virial radius but rapidly cools down to below $3\times10^4$~K.   The gas that ends up in this central zone, for the most part, remains there and even when feedback dramatically increases the temperature of some of the particles, their evolution is more akin to a fountain.  They cool down very rapidly and return to the central zone.   %On the whole, these particles do not migrate outward.

The middle zone (second panel in each column) is built by circulation flows. 
At early times, the gas enters the virial radius through cold flows and ends up in the galaxy.   Subsequently, it is heated by SNe-driven or merger-induced shocks and expelled from the galaxy.   Some of the gas directly contributes to the hot and warm-hot gas in the second zone, while the rest flows out of the halo and then, falls back in and is shock-heated upon re-accretion.  Around $z=1.5$, the pattern changes. The incoming gas follows two different paths in the $r-T$ diagram: some of the gas follows the trajectory of the gas that ends up in zone 1, except that this gas is subsequently heated and expelled from inner halo.   While the rest is heated via accretion shocks to $T>10^5$~K and flows directly into the middle zone.   That both hot and cold mode accretion exist at the same time is expected while the halos are transitioning from the predominantly cold to predominantly hot \citep{Keres:2009,Voort:2011}.   
{\it A significant fraction of the gas that comprises the hot corona is that which has been heated, and often driven outward, by shocks occurring within the galaxy or deep in the inner halo.}
%obtains SNe energy or shock-heats because of mergers, and is ejected outward, even beyond the virial radius. Then in the second stage, the trajectory of the incoming particles forks in the 

Finally, the gas in the outermost zone (third panel) initially comes in primarily as cold flows, is heated and expelled from the galaxy, and expands to fill the outer zone. Some of this gas expands beyond the virial radius and eventually falls back in.  While expanding, the gas cools and drops below the $10^6$ K threshold.  This is the clearest indication that a fraction of the outer envelope of warm-hot gas in fact started out as hot gas.  Around $z=1.9$, this inflow-to-outflow channel is overtaken by a different pathway: the gas is heated just prior to crossing the virial radius. This gas receives no SNe energy; accretion shock is the primary heating agent.   However, a fraction of the accreting gas still cools, flows into the galaxy, and returns as outflow.   
}

\begin{figure*}
\includegraphics[scale=0.5]{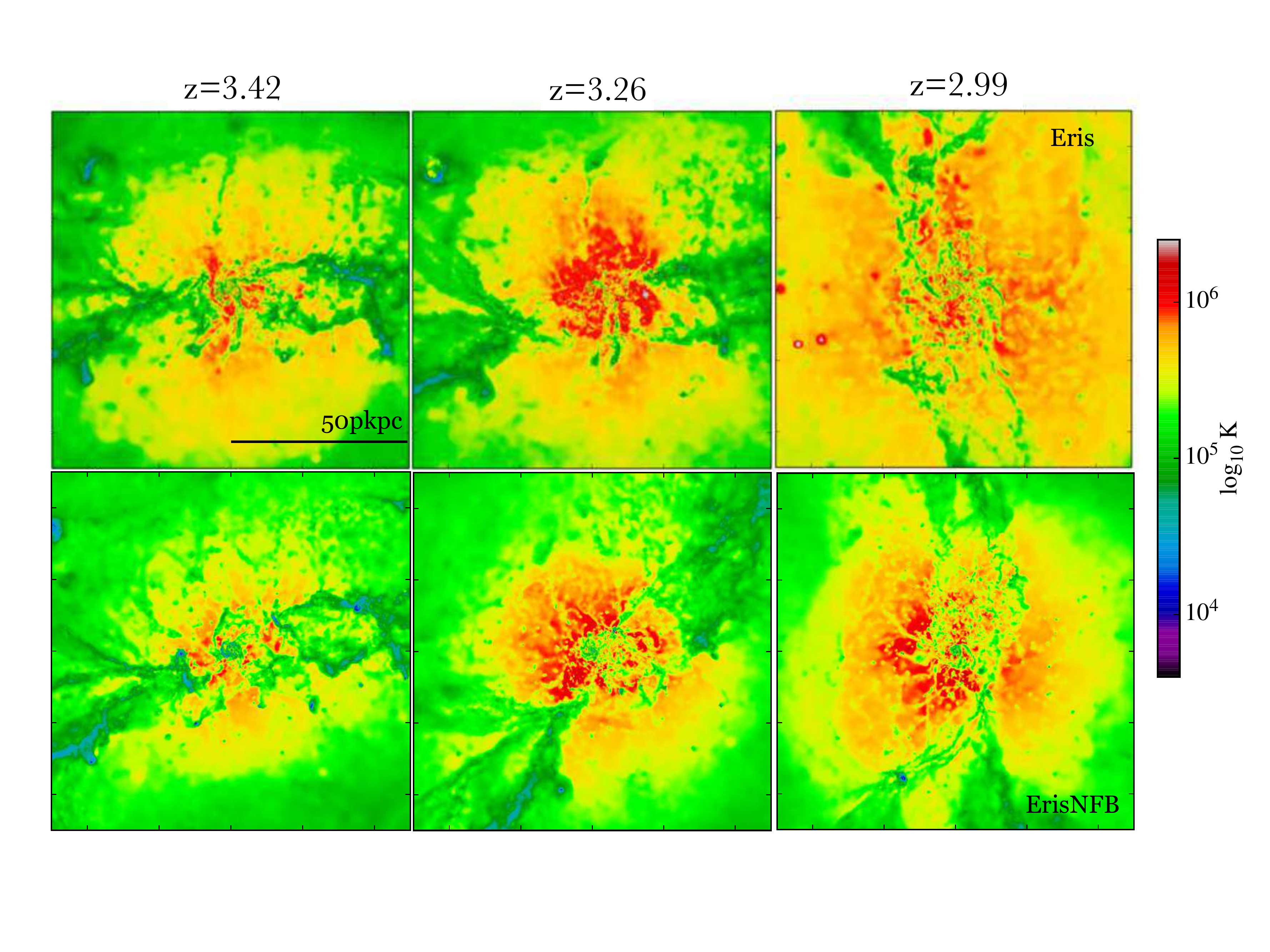}
\centering
\caption{Temperature maps of a gaseous halo at three timesteps, illustrating a typical ``blast" in action in case with feedback (\eris, top row) and without (\erisnfb, bottom row).  There is a lot more hot (red) and warm-hot (light green-yellow-orange) in the simulation with feedback than without, and this gas extends further out.
The width of each square is 100~pkpc.}
\label{fig:blast}
\end{figure*}

Next, we quantify the difference that the presence of SN feedback makes on the diffuse gas budget and its evolution.  In order to do so, we perform a direct test of the impact of SN feedback on the halo; namely, we run another simulation (hereafter, \erisnfb), which is identical to \eris~in every aspect except that the  SN feedback is switched off after $z=4$.  This redshift was chosen as a compromise involving various competing concerns:  (i) the simulation completes in a reasonable period of time;\footnote{Simulations that allow for cooling but no feedback tend to slow down dramatically as cold gas builds up.} (ii) the diffuse gaseous halo should ideally be absent; and (iii) the lack of feedback should have little or no influence on the assembly history.
Although one cannot produce realistic galaxies without feedback, the benefit of this test lies in being able to isolate the effect of SN heating from gravitational heating by accretion shocks and merger-induced shocks.  We note that although we do the above experiment with \eris, we expect similar results for \venus.

{ Figure~\ref{fig:blast} illustrates a violent heating episode, which is an example of a series of such events occurring repeatedly over the lifetime of our \eris~run, and compares the magnitude of this event in the original \eris~run with feedback (top row) and in \erisnfb~ (bottom row).}  The temperature maps in Figure~\ref{fig:blast} capture the evolution of the ``blast"  {over 0.3~Gyr}.  {In both \eris~ and \erisnfb, the general sequence is similar:} at $z=3.42$, the gaseous halo consists of two bubbles of warm-hot gas that are separated by inflows of cold/warm gas. By $z=3.26$, the central concentration of over one million-degree gas grows significantly into a nearly spherical region approximately 20~kpc in radius.  Thereafter, the hot gas expands and cools.  Early ``blasts" in \eris~are easy to detect in the temperature maps, and their timing coincides with peaks in the star formation rate.
%\citep[see Figure~\ref{fig:SFHr80} in the Appendix, and also][]{Voort:2016}. 

The second row of Figure~\ref{fig:blast} shows that the hot gas can also be generated much earlier than expected even in the absence of feedback.  {As early as $z=3.42$, gas in the center of \erisnfb~exceeds the temperature threshold of $10^6$~K. Here, the outflows are generated by {shock-heating associated with a major merger} and can seen in the map as conical warm-hot and hot patches extending beyond the virial radius of the halo at that epoch.}  However, the outflows are weaker and contain less hot gas (by mass) than those generated in \eris, where Type II supernovae feedback acts in concert with the merger.  The extent to which the heated gas expands by $z=2.99$ in \erisnfb~compared to \eris~(bottom right panel to the top right panel in Figure~\ref{fig:blast}) attests to this.  

%The metal-poor outflows generated without feedback are triggered by  {shock-heating associated with mergers}. However, those outflows are much less powerful than those combined with SN feedback, as they have a limited range compared with the heat maps of \eris at $z=2.99$. This result indicates that SN feedback has a significant impact on the gas distribution of a gaseous halo even as early as $z=3.5$.

\begin{figure}
\hspace{-1cm}
\includegraphics[scale=0.65]{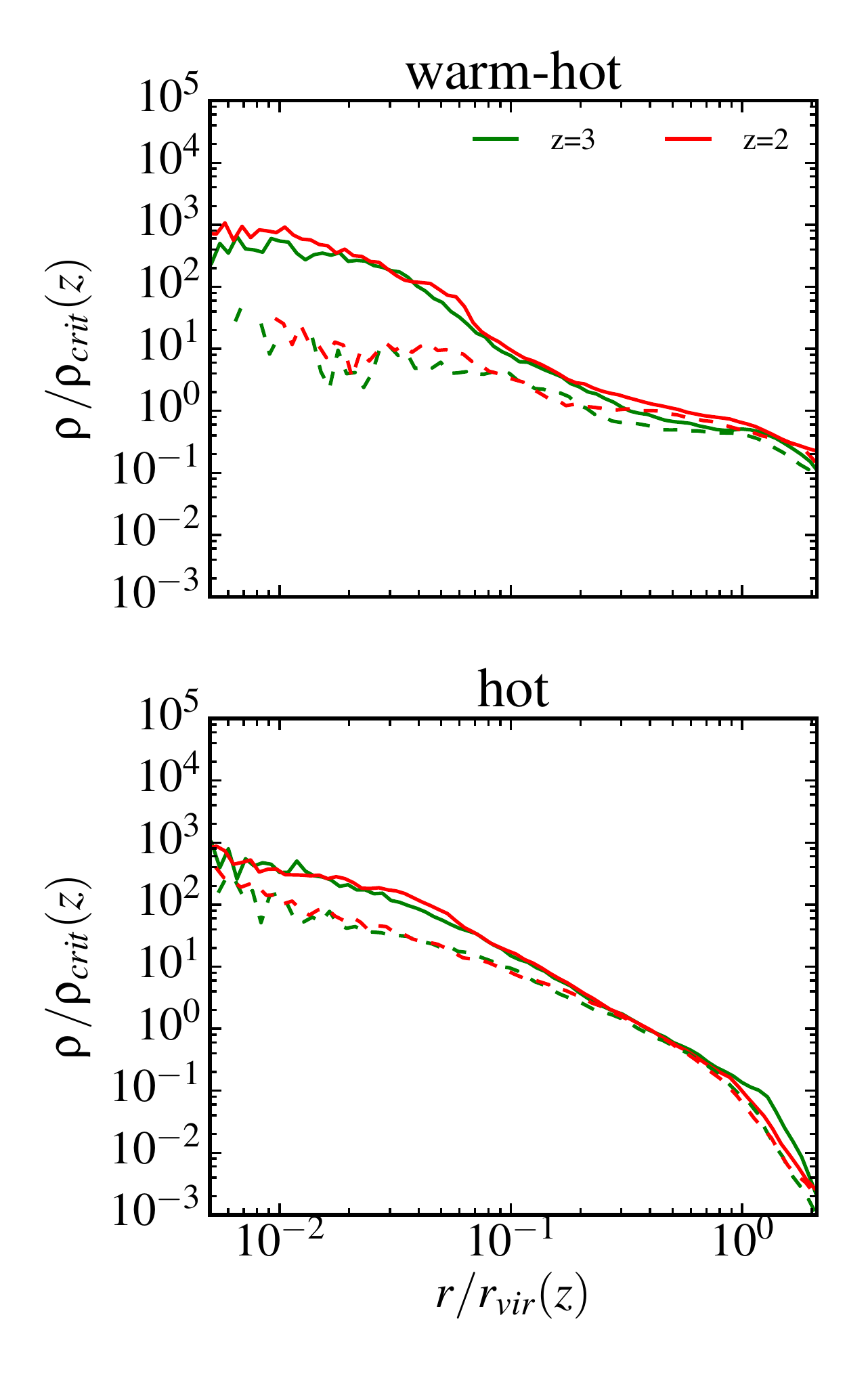}
\centering
\caption{Density profiles of warm-hot (top) and hot (bottom) with SN feedback (solid lines) and without (dashed lines). The density and radius are normalized to the critical density of the universe and virial radii of halos at the redshifts under consideration.}
\label{fig:FBNFB}
\end{figure}

 {In Figure~\ref{fig:FBNFB}, we show the radial density profiles of the two gas phases: warm-hot (top) and hot (bottom) for the run with SNe feedback (solid lines) and without (dashed lines) at $z=3$ and $z=2$. The absence of feedback results in significant differences in the inner region of the diffuse gaseous halo ($ \lesssim 0.3 r_{vir}$), particularly with respect to the warm-hot phase. The density of the warm-hot gas in that region is almost two orders of magnitude higher if SN feedback is switched on.  The difference between the density of hot gas in \eris~ and \erisnfb~ is slightly smaller but no less significant, namely hot gas is about five times denser in \eris~ than in \erisnfb.  By $ \sim 0.6 r_{vir}$, the density profiles of both gas phases converge.}  In terms of mass, we find 1.3 times more hot gas in \eris~ than in \erisnfb~ within the virial radius at $z=3$ and by $z=1$, that ratio increases slightly to 1.5.  In total, introducing feedback increased the abundance of (warm-hot + hot) $T>10^5$~K gas within the virial radius by a factor of 1.3-1.5 between $z=3-1$.

The gaseous halos in \eris~ and \erisnfb~ are not only structurally different but also result in divergent X-ray evolution with redshift.  We compute X-ray luminosities at $z=3,\;2,\;1$ in the $0.5-2$~keV band for \eris~ and \erisnfb~ using the same procedures as described in the start of Section~\ref{sec:results}.  {We find that the} X-ray luminosities are higher when feedback is included, namely $3.40\times 10^{41}$, 
$8.9\times 10^{40}$, $5.9\times 10^{40}$ erg/s as opposed to $2.4 \times 10^{38}$, $1.1 \times 10^{38}$, $ 1.3 \times 10^{37}$ erg/s at $z=3,\; 2,\; 1$ respectively.  {\it In the absence of feedback, the gaseous halo is considerably under-luminous and in fact, fails to match the observed X-ray luminosity.}

\arif{
The above result is the product of three effects.  In the no-feedback run, amount of hot gas expelled out of the galaxy is lower to start with, and this gas tends to cool down faster.  Both of these lead to a smaller, less massive corona and a lower X-ray luminosity \citep[see Figure \ref{fig:blast} as well as ][]{Toft:2002}.  Additionally, as illustrated in Figure \ref{fig:FBNFB}, the run with SNe feedback has a considerably higher density of hot (and warm-hot gas) within the inner $\sim 0.6 r_{vir}$.  Since the X-ray emissivity scales as 
$\rho_{gas}^2$, this makes for a significantly brighter inner halo. 
%may seem counter-intuitive at first glance but in the absence of feedback, too much of the halo gas cools out.  The reduced X-ray luminosity is the result of there not being enough hot gas in the halo.   Over-cooling also results in too many stars in the galaxy as well as changes in the structural properties of the disk.   To return to the differences in gas mass discussed above, feedback not only drives gas out of the galaxy and into the halo, it also heats the gaseous halo and suppresses cooling flows.
SN feedback is, therefore, not only essential for obtaining realistic disk galaxies but also realistic diffuse halos.}

\arifho{
Before continuing, we would be remiss if we did not point that our findings about the importance and impact of heated galactic outflows on the formation and evolution of the gaseous halos is at odds with the assertions of \citet{Fielding:2017}.  There are a number of fundamental differences between our and their studies that may account for the divergent findings, the most important of which is that  the \citet{Fielding:2017} study is based on simulations in which the galactic halos are represented by initially spherically symmetric gas profiles designed to capture the low-redshift structure of the systems in an idealized fashion.  And, although they introduce effects that, over time, break the strict spherical symmetry,  the absence of the proper time-dependent cosmological framework for how gas accretion proceeds means that the simulations do not allow for strongly non-linear spatial asymmetries in the form of mergers and filamentary accretion that arise in and are key features of realistic hierarchical cosmic structure formation simulations, nor do they treat the important effects like the transition from cold to hot mode, etc.  As Figure \ref{fig:HC} shows all of these, the different accretion modes and their associated geometries, the mergers, and the galactic outflows triggered by both mergers and stellar feedback, play a pivotal role in the formation and evolution of MW-like halos and corresponding symphony (or competition) cannot be neglected even at $z=0$.
}

\subsection{Corona and the cold flows}\label{sec:bimodality}

As discussed in the introduction, the galactic halos of MW-like systems 
occupy a transition mass scale, prior to which the halos are expected to  {acquire gas primarily via the ``cold mode" and thereafter via the ``hot mode".}  In this picture, the gas that enters the halo at early times, comes in via  cold, dense filamentary streams, remains cold, and ends up flowing onto the central galaxy.  As we have demonstrated in this paper, some of this gas is expelled from the galaxy as a result of supernovae-powered and merger-induced shock-heating, thus initiating the inside-out formation of an atmosphere of hot and warm-hot gas in the halo.  This happens at an earlier epoch than what was previously established, and in halos of lower mass  than the critical mass of $\sim 10^{12}M_{\odot}$\arif{, i.e. before the gaseous halo can be established by purely gravitational means.}  The emergence of this atmosphere sets the stage for a stable accretion shock at the virial radius.  Concurrently, on the supra-galactic scale, the cosmic filaments are growing and as their widths become comparable to the sizes of the halos, an increasing fraction of the gas flows into the halos in quasi-spherical fashion, shock heating as it encounters the accretion shock. 
%At some point, the cooling efficiency of the inflowing gas in the filaments declines, the cooling time exceeds the compressive heating time, and the gas diverts from the galaxy and also contributes to the galaxy's diffuse gaseous halo \citep{Joung:2012}.  
 
There is one outstanding question that we have touched upon previously but haven't yet examined and that is, whether the diffuse gas in the halos and specifically the outflowing gas expelled from the galaxy affects the galaxy-filaments connection. We briefly discuss this connection in this subsection.  A detailed analysis will be presented in a follow-up paper.

\begin{figure*}
\centering
\hspace{0.1cm}
\includegraphics[scale=.27]{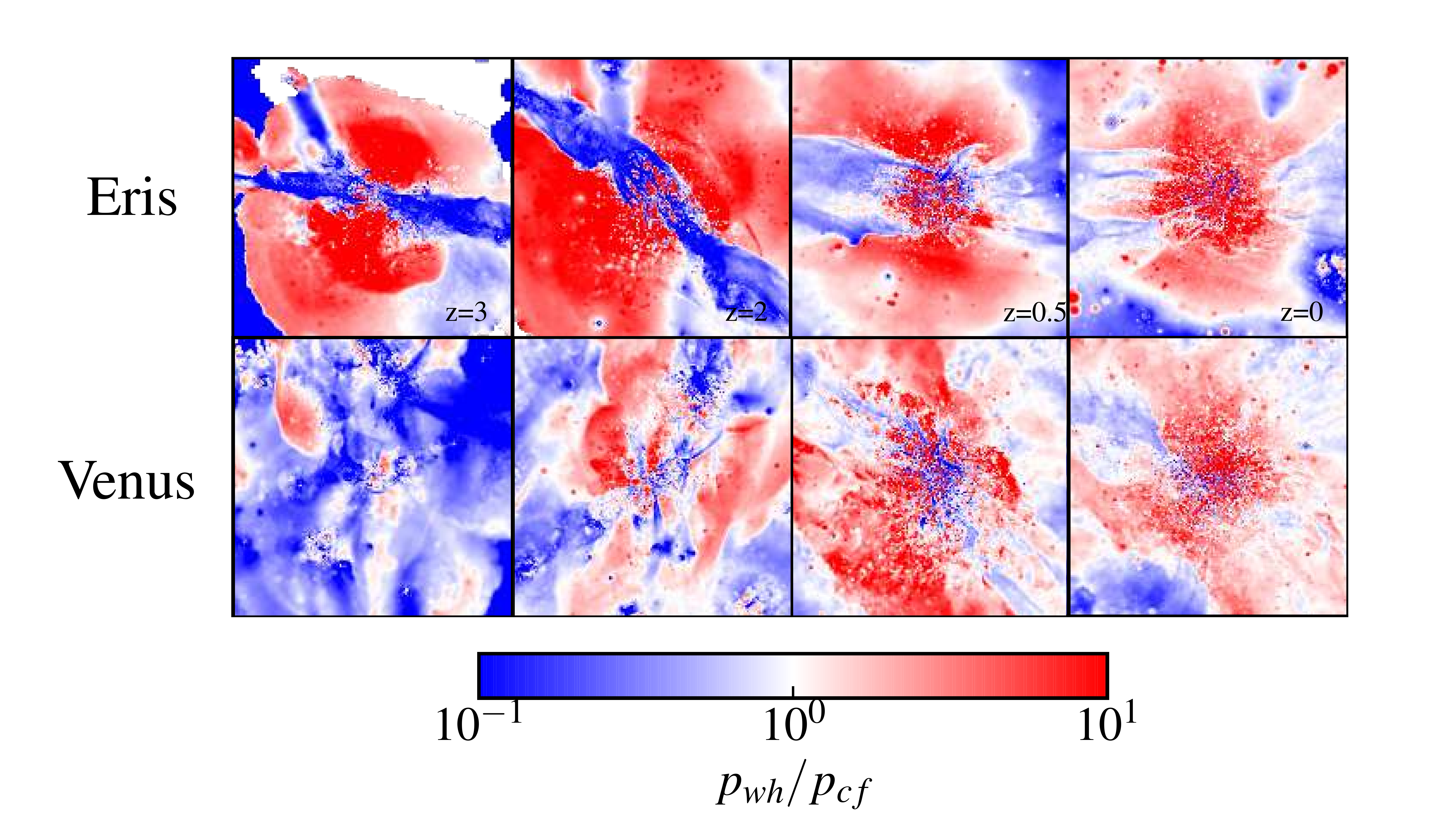}
\caption{Snapshots of the ratio of the total pressure (i.e. the sum of ram and thermal pressure) of the $>10^5$ K gas to the total pressure of the $<10^5$ K inflowing gas, at two redshifts when the cold mode accretion is dominant (two left panels) and at two redshifts when the hot mode acceleration is ascendant (two right panels).  The width of each square is 1~pMpc.}
\label{fig:HC}
\end{figure*}

\arif{
During explosive outflow events while the halo is in the cold mode accretion phase, the gaseous halo can be thought of as comprising two colliding, shearing  and sometimes turbulent flows: cold dense inflowing filamentary streams and powerful hot/warm-hot outflowing wind.  The wind generally follows the path of least resistance and streams around and often perpendicular to the filaments. However, in the inner halo, two opposing flows do interact strongly and exchange energy and momentum.   

To illustrate what happens at these times, we show in Figure~\ref{fig:HC} snapshots of $p_{wh}/p_{cf}$ at four different redshifts.  Here, $p_{wh}$ is the total pressure (i.e., the sum of thermal pressure, $p_{th} = \rho kT / \mu m_p$, and ram pressure, $p_{ram} = \rho v_r^2 /2$) of the $T > 10^5$~K gas, and $p_{cf}$ is the total pressure of the cold inflowing gas, defined as gas with temperatures $T < 10^5$~K and $\vec{v}\cdot\vec{r}<0$.  In Figure~\ref{fig:HC}, blue regions correspond to the inflowing streams and red to hot outflowing wind streaming out of the galaxy.  

In the absence of powerful outflows, the disk in \eris~is fed by two dominant, oppositely oriented, streams that drill their way to the center of the halo and connect smoothly to the disk.   In \venus, the disk is bombarded by clumps of cold gas from all directions, reflecting the active merging history of this run.  Eventually, the inflow in \venus~settles down into the same pattern as \eris; however, this happens at $z \approx 1.5$ so we will concentrate on \eris.  

At early times (i.e. $z \gtrsim 2$), the outflows have a relatively limited impact on the filaments.  Looking closely at the center of the $z=3$ and $2$ panels, we observe that the blue streams, which ordinarily would converge and terminate on the central galaxy, have been disrupted. The ram pressure of the outflowing wind causes the mouths of the streams to broaden and break-up into thin rivulets.  However, this delta-like feature remains relatively close to the disk. 
And, once the winds wane, the steady inflow of cold gas along the streams re-establish the filaments within the inner halo, which then reattach to the central galaxy.  However, as the filamentary flow begins to weaken, the total pressure of the outflows are not only able to cause the streams' mouths to fray and broaden, but disconnect them from the galaxy and push the mouths away to increasingly larger radii.  In effect, the winds accelerate the destruction of the streams. 
}

\begin{figure*}
\centering
\includegraphics[scale=0.25]{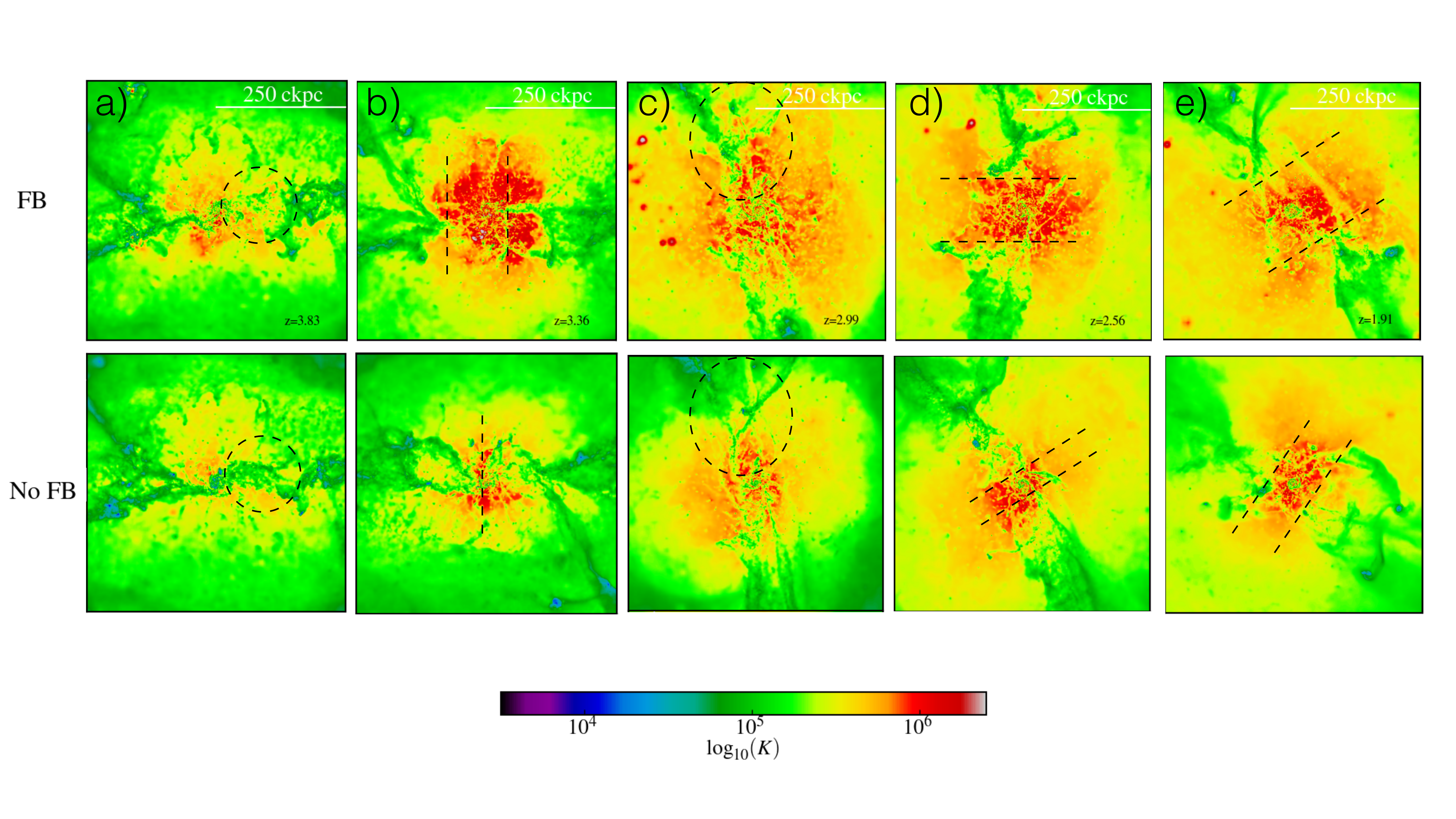}
\caption{Temperature maps of \eris~ (top row) and \erisnfb~ (bottom row) between $z\sim 4$ and $z\sim 2$ comparing the change in the location (lines, columns b, d and e) and structure (circles, columns a and c) of the filaments with and without feedback.}
\label{fig:filamFBNFB}
\end{figure*}

%\arifhum{We showed in section~3.2 that inclusion of SN feedback increases the density of diffuse gas near the galaxy by up to two orders of magnitude. Hereafter, we describe qualitatively the consequences of efficient SN feedback for the galaxy-cold flow connection. Figure~\ref{fig:filamFBNFB} presents the temperature maps of gas in a face-on view on the central galaxy.  Top row shows a few early snapshots of the \eris simulation and is set against the snapshots of \erisnfb in the bottom row in order to show the structural differences in the halo at the absence of feedback. 

\arif{
In Figure~\ref{fig:filamFBNFB}, we compare qualitatively the location and the structure of the filaments in \eris~ and \erisnfb~ at five snapshots at redshifts $2 \lesssim z \lesssim 4$ to gauge specifically the impact of SNe-powered outflows.  The panels show the temperature maps of the gas in a face-on view of the central galaxy, with \eris~results presented in the top row and \erisnfb~results from the same epochs juxtaposed in the bottom row.  The width of each square is 500 comoving Mpc.

Looking first at \eris~ results (top row), we note that intially the cold streams easily reach and feed the galactic disk of \eris~(panel a).  Powerful outflows fray the mouths of the streams and push them back, often detaching them for extended periods of time from the galaxy
(in panel b, the push-back region is marked by the two dashed lines).  Frequently, the filaments manage to maintain connection with the galaxies via many small rivulets (c) but as the coherence of the streams progressively weakens, they are more easily detached from the galaxy (d, e).   

The outflows in \erisnfb~(lower panels) act somewhat similarly but in this instance, are merger-related and are much weaker. Consequently, they don't affect the filaments as strongly as in \eris.  Comparing the top and bottom panels in columns b, d and e, we find that the outflows engendered by mergers and feedback acting in concert (top row) are able to push the mouths of the filaments back further.  Also, the merger+feedback-fed outflows and the coronal gas start to heat and strip the filaments as early as $z=3.83$ (see circles in columns a and c).

In effect, feedback not only helps maintain a larger mass of gas in the hot corona and the warm-hot atmosphere, it erodes the cold flows more effectively and accelerates the breakdown of the direct connection between the streams and the central galaxy.  This in turn contributes to the regulation and quenching of star formation in the central galaxies.  In effect, our results suggest that although SNe-powered outflows are categorized as ejective feedback (i.e. star formation is quenched by expelling the fuel for star formation), in practice they also act to strangle the galaxy by choking off its supply of fresh gas.  In other words, SNe-powered outflows act both as ejective \emph{and} preventive feedback.   }

\arif{
The above result raises questions about the kinetic ``decoupled winds'' approach \citep[e.g.][]{Springel:2003,Oppenheimer:2006,Oppenheimer:2008a,Illustris:2013,Liang:2016,Dave:2016,Illustris:2018} to modeling stellar feedback.  As implemented at the present, this approach is not designed to act in a preventative capacity along the lines described above. As such, one would expect that the feedback scheme likely needs to be super-ejective to achieve the same results.   We will return to this issue in future work.}

%\arifho{
%In this section, we provide a broader context to our primary findings, as described in Section 3 above, by briefly comparing our results to other recent relevant simulation studies, including the \eiik~run.}

\subsection{\eiik}\label{sec:e2k}

\arif{
As mentioned in Section 2,  \eris~and~\erisnfb~ are part of a larger suite of simulations carried out using identical initial conditions where we experiment with the modeling details of the various sub-grid physical phenomena, including star formation, cooling and feedback, in order to see if we can successfully reproduce the observed stellar mass--halo mass relationship across a range of redshifts \citep{Behroozi:2013,Moster:2018} while still ending up with a late-time spiral galaxy that matches the Milky Way as well as \eris.  Here, we briefly discuss the \eiik~simulation.  As described in Section 2, the key defining features of \eiik~are (i) metal diffusion, (ii) the introduction of the metal-line cooling channel for the $T > 10^4$ K gas, (ii) more efficient cooling of $T < 10^4$ K gas, and (iv) boosted SNe feedback \citep{Shen:2012,Shen:2013,Sokolowska:2016,Sokolowska:2017}.  

Metal-line cooling in \eiik~is computed via tables generated using {\sc cloudy} \citep{cloudy} under the assumption that the metals are in ionization equilibrium \citep{Shen:2009aa} in the presence of an updated cosmic ionizing background \citep{Haardt:2012aa}.  One could argue that the lack of metal-line cooling in \eris~is not self-consistent since the very SNe-powered outflows that appear to play a crucial role in the early assembly and evolution of the gaseous corona also continuously introduce metals into the warm-hot and hot CGM.  A number of studies (e.g. \citealt{Voort:2011}) have shown that the increased efficiency of cooling associated with metal-line radiation results not only in cooler temperatures for the halo gas but also in an increased deposition of gas onto the galaxy. In fact, in the absence of any counteracting mechanism, metal-line cooling can lead to unrealistically high stellar mass.

Additionally, as mentioned in Section 3.1, enhanced cooling also leads to a higher halo mass threshold at which the accretion onto the halos transition from predominantly cold to predominantly hot mode (and can sustain a stable virial shock; c.f.~Figure~\ref{fig:form_vs_mcrit}).  \citet{Voort:2011}, who allow for both self-consistent enrichment of the gas   and metal cooling in their simulations, find the threshold mass ranges from $1.6\times 10^{12}\;M_\odot$ at $z\approx 3$ to  $1\times 10^{12}\;M_\odot$ at $z\approx 2$, to $7\times 10^{11}\;M_\odot$ at $z\approx 0$.  Given their identical mass accretion history, the \eris~suite of simulations (i.e.~\eris, \erisnfb, \eiik, etc.) cross this higher mass threshold at $z\approx 0.3$.
}

The recipes for star formation and SNe feedback in \eiik~are the same as in \eris~and~\erisnfb; however, there are some differences in the values of some of the controlling parameters.  For instance, the star formation threshold $n_{\rm SF}$ in \eiik~is set at 100 atoms~cm$^{-3}$; the maximum temperature of a particle allowed to participate in star formation is $T_{\rm max} = 1\times 10^4$~K; supernova efficiency parameter is increased to $\epsilon_{\rm SN} = 1.0$; and the stellar initial mass function is updated to \citet{Kroupa:2001aa} (see Table~\ref{tab:tab0} in Section 2 for an overview of the differences between \eris~and~\eiik).

The strength of feedback depends on the number of SNe produced, which in turn is governed globally by the IMF and locally by the star formation density threshold. The revised IMF yields about a factor of 2.8 more SNe for the same star formation rate. Moreover, as explained in detail in \citet{Guedes:2011aa} and \citet{Mayer:2012}, the local star formation rate, and thus the local impact of SNe, can be boosted significantly by raising the star formation density threshold and allowing the ISM to become more inhomogeneous, an effect that saturates only at very high resolution and density thresholds, well above those resolvable in the current generation of  cosmological simulations \citep{Hopkins_et_al_2012}.  

In \eiik, SNe feedback is boosted both globally and locally using parameters that were chosen to achieve realistic stellar masses in accordance with abundance matching at high and low  redshift.  And indeed, owing to a more effective squelching of star formation at early times, the simulation yields a much better agreement to the \citet{Behroozi:2013} stellar mass--halo mass relationship than \eris~and its other variants up to $z\sim 0.5$.  The stellar mass at earlier epochs is lower than in \eris.   At low redshifts, however,  \eiik~deviates significantly from \eris~and the corresponding galactic system is a much less faithful replica of a Milky Way-like late type spiral \citep{Sokolowska:2017}.

\begin{figure}
\centering
\includegraphics[scale=0.4]{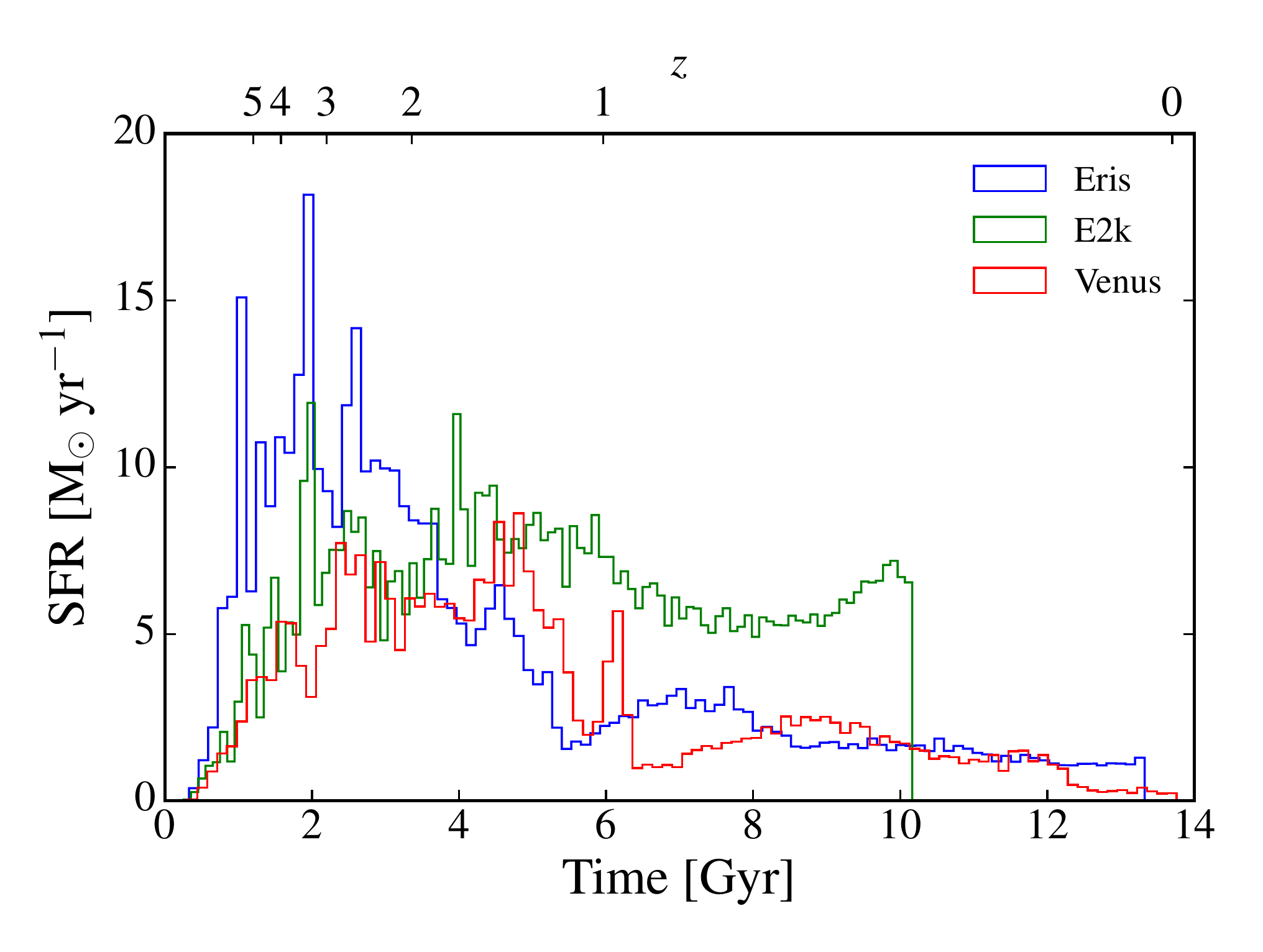}
\includegraphics[scale=0.57]{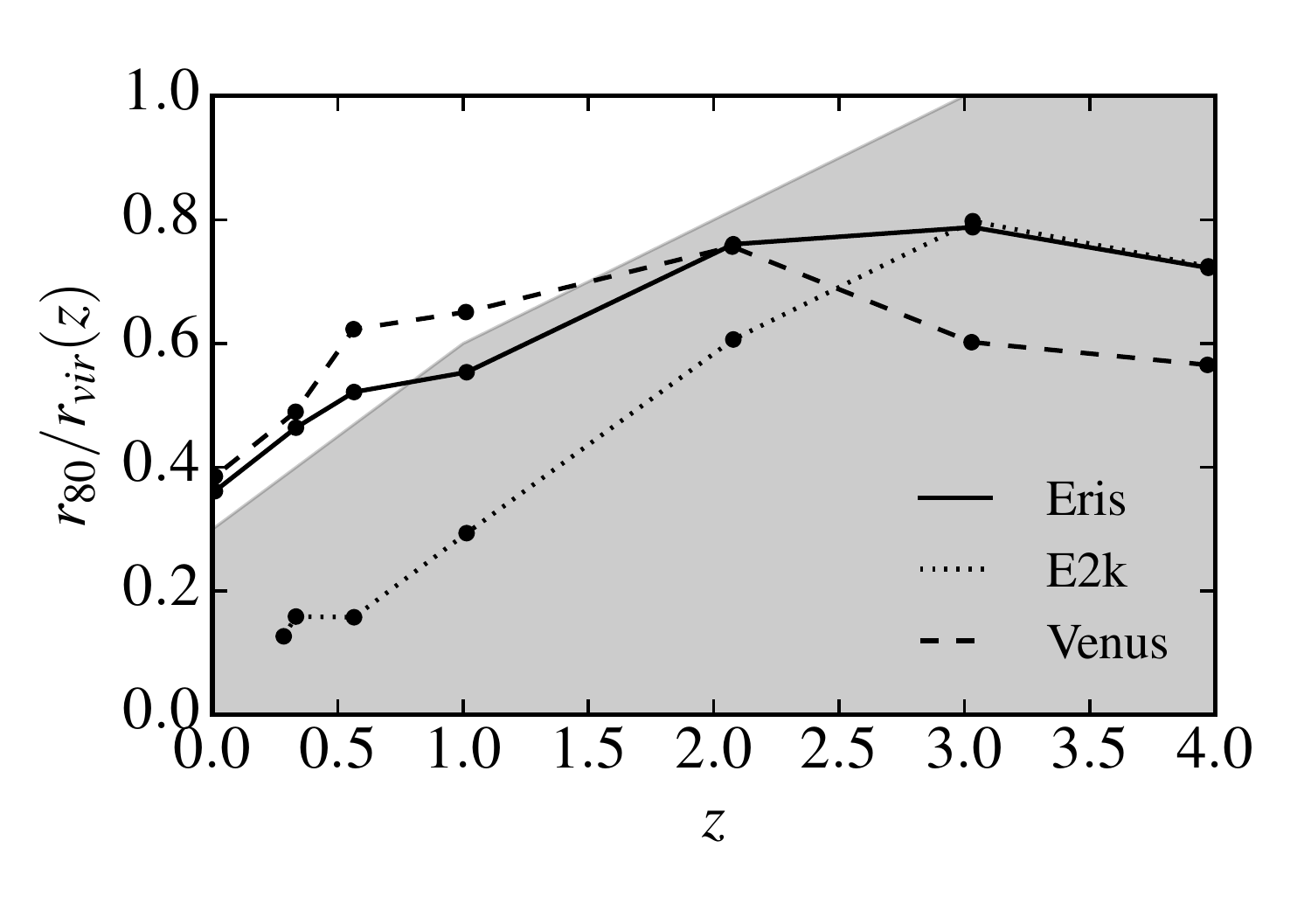}

\caption{Top panel. The comparison of star formation histories of \eris, \eiik and \venus. Note that the run \eiik was stopped after 10~Gyrs of evolution. Bottom panel. The radius of a hot corona encompassing 80\% of its mass as a function of redshift.}
\label{fig:SFHr80}
\end{figure}

In the top panel of Figure~\ref{fig:SFHr80}, we juxtapose the star formation histories of \eris, \venus, and \eiik. \eris~and \venus~differ at early times. \eris~has a much higher star formation rate until $z=2$; thereafter, it drops steadily over the course of 2 Gyrs and stabilizes at $\sim 1-2\;{\rm M}_\odot/{\rm yr}$.   The star formation in \venus~rises more slowly and reaches the peak value later compared to \eris, and the epoch of high star formation is slightly longer.   This is due to \venus's late assembly.  At late times, however, the star formation settles at a similar rate to that in \eris.

The star formation history in \eiik~differs from \eris~at all epochs and in this case, this is due to differences in how the subgrid physics unfolds in the two simulations.   The boosted feedback in \eiik~ successfully tempers the star formation rate at early times and allows the system to evolve in sync with abundance matching expectations by expelling more gas out of the halo and pushing this gas further away than in \eris~(see Table 3 in Paper I: the fraction of baryons within the virial radius ($r_{vir}$) and $3r_{vir}$ is 71\% and 91\% in \eris, compared to 67\% and 74\% in \eiik).   This balance between cooling, star formation and feedback works well until approximately $z\approx 0.6$, after which it breaks down.  The forming galaxy's deepening potential well increasingly confines the galactic wind within the halo. In the case of \eiik, the metal-line radiative losses accelerate the cooling of the CGM and the resulting increased flow of gas onto the galaxy provides fuel for enhanced star formation.  Consequently, the already high star formation rate (compared to both \eris~and the Milky Way galaxy) begins to rise. Overall, \eiik~ converts more gas into stars and this difference is largely due to late-time star formation:  At $z=0.5$, the stellar masses in \eris~and \eiik~differ by only 1\%.  By $z=0.3$, \eiik~has 10\% more mass in stars: $M_* = 3.6 \times 10^{10}~M_{\odot}$ and $4\times 10^{10}M_{\odot}$ in \eris~and \eiik, respectively. And given the rising star formation rate in \eiik, we expect the difference to continue to grow to $z=0$.\footnote{We note that the \eiik~simulation had to be stopped at $10$ Gyrs due to the exhaustion of computational resources.}  Additionally, the galaxy in \eiik~ has a kinematically hotter disk \citep{Sokolowska:2017} and as discussed in Paper I, the X--ray luminosity of the \eiik~system in the 0.5-2~keV band is $L_X =3\times 10^{42}$ erg/s, which is a 100 times higher than the \eris~result as well as the most current estimates for the Milky Way system.  

The bottom panel of Figure~\ref{fig:SFHr80} illustrates one of the reasons why the X--ray luminosity of \eiik~is some much higher.  The figure shows the evolution of the radius encompassing 80\% of a hot corona in \eris, \venus, and \eiik~(c.f.  Figure \ref{fig:r80}).  Prior to $z=3$, $r_{80}$ in \eris~and \eiik~evolve in lock-step but after $z=3$, the $r_{80}$ in \eiik~ begins to shrink and drops to half of the value of $r_{80}$ in \eris.  The reduced size of the corona in \eiik~is not due to there being less mass. The amount of gas in the corona at $z=0.5$ in the two simulations differs by only 10\%, with the corona in \eiik~being the more massive of the two \citep{Sokolowska:2016}.  The coronal gas distribution in \eiik~is more compact and the mean gas density is a factor of $\sim$10 higher.  In addition, as reported in Paper I, the hot corona of \eiik~has significantly higher metallicity that, even at comparable densities, results in increased X-ray luminosity via line emissions as well as  boosted bremsstrahlung radiation. The combination of higher density
and higher metallicity explains why the X--ray luminosity of the system is a factor of 100 larger than in the other simulations.  The enhanced cooling also results in more robust, thicker cool-gas filaments that are more resilient to disruption.

That \eiik~fails to match the properties of the MW system (i.e. the galaxy and its gaseous halo) highlights potential concerns with the current treatment of metal-line cooling \citep[c.f.][]{Christensen:2014,Tremmel:2018}.  One possible problem is that metal-line cooling is computed under the assumption that the ions are in ionization equilibrium, which can lead to an overestimate of the cooling rate if the ions are not in equilibrium. \citep{Oppenheimer:2013} show the warm-hot component of the CGM is most susceptible to this and the effect is even stronger if the gas is subject to fluctuating ionizing radiation from an active galactic nucleus \citep{Oppenheimer:2013a}. It is also highly likely that the lack of heating sources, such as  compact X--ray binaries \citep{MadauFragos:2017}, cosmic ray heating, photo-electric heating, etc. in our simulations may also be a contributing factor.  Also, we do not account for the effects of local ionization radiation, which can also alter the cooling profile of the CGM.  Additionally, \citet{Christensen:2014} --- see also \citet{Tremmel:2018} --- have shown that the inclusion of metal cooling in simulations that do not have sufficiently high resolution to support the proper modeling of molecular hydrogen physics and the multi-phase structure of the ISM and the CGM results in over-cooling.  In simulations that attempt to do so, they find that the resulting galaxies have star formation histories and outflow rates that are more similar to those in the primordial cooling runs (like our \eris) than to those that include metal line cooling but have inadequate resolution (like \eiik). 

In summary, the combination of boosted feedback and metal cooling, particularly, has a significant impact on both the evolution of resulting galactic system, including the structure of the galactic disk and the corona, and the ability of SNe-driven shocks to disrupt the cold filaments feeding the galaxies. However,
given the various concerns about the modeling of metal cooling in simulations at the present time,  it is not clear whether \eiik~ is necessarily a better representation of reality.  In fact, running without metal cooling may well be preferable as the approach appears to better approximate the dynamics and thermodynamics of gas components with moderate to high average densities. 

\section{Conclusions} \label{sec:conclusions}

\arifho{
We use a set of 4 hydrodynamical zoom-in simulations of Milky Way-like galactic systems --- described in Section 2 --- to study the halo-filling gaseous atmosphere cocooning the central galaxy.  We analyzed the various properties of the gaseous halos in our simulated systems in Paper I \citep{Sokolowska:2016}, and compared these against available observational results for the Milky Way.  In the present paper, we investigate the mechanisms that contribute to the formation of the gaseous halo. 

%\arifha{Briefly, the runs include: (i) The \eris~simulation, which not produces a remarkably realistic Milky Way-like galaxy \citep{Guedes:2011aa} but the properties of this virtual MW's gaseous halo are also in excellent agreement with recent observational constraints for our Galaxy's halo \citep{Sokolowska:2016}. (ii) The \venus~run, which is identical to \eris~in terms of the physics but uses different initial conditions that results in a much more active merger history down to fairly low redshifts (in contrast to \eris's relatively quiescent low redshift merger history) and delayed assembly of the final halo.  (iii) \erisnfb, which is essentially \eris~restarted at $z=4$ and continued forward without SNe feedback in order to isolate the effects of accretion shocks and merger-induced galactic outflows from those of feedback.  And, (iv) \eiik, a run in which $T > 10^4$ K is subjected to metal-line cooling and SNe feedback is boosted. Neither the final galaxy nor its gaseous halo in \eiik~present a good match MW system but the system does track the stellar-to-halo mass relation at both high and low-z much better than any of the other runs.}
   
The conventional view is that the gaseous atmosphere arises when the cosmological influx of gas onto a galactic halo transitions to diffuse quasi-spherical inflow.  This gas is heated to near-virial temperatures by accretion shocks and in the interior of the halos, is heated further due to compression as long as radiative cooling is relatively inefficient.  While this picture is valid in the sense that the late time mass budget of the gaseous halo is generally dominated by gas accreted in this fashion, the details of how the gaseous halos arise are considerable more involved.  We find, and this is our most important result, that \emph{the gaseous halo, and especially the corona,\footnote{We categorize the gas comprising the gaseous halo into two components: the hot ($T>10^6$~K) X-ray luminous corona and the dominant (in terms of mass) warm-hot ($T=10^{5-6}$~K) component.} are not simply the products of hot spherical accretion.  We trace their origin to gas expelled from the galaxy by mergers-induced shock heating and SN feedback.  This is why they are present in our simulations much earlier than expected (as early as $z\approx 3-4$) and can be seen even  around lower mass (e.g.~$M_{\rm halo} \lesssim 10^{11}\;M_{\odot}$) halos. }  

In detail, our simulations show that prior to $z\approx 2$, most of gas comprising the halo-filling warm-hot atmosphere is gas that was expelled from the central galaxy and cooled as it expanded outward.  Only after $z\approx 2$ does this inside-out process become inverted. Moreover, we find that direct SNe heating is the dominant heating mode for the warm-hot gas at all redshifts. 

The corona, which comprises gas with temperatures a factor of $2-2.5$ greater than virial temperature of the system at $z=0$, is of particular interest because although it is not the most abundant component in terms of mass, it is the phase that is primarily responsible for the diffuse X-ray emission and for OVII/OVIII emission/absorption observed in the halo.  In our simulations, the coronal gas is centrally concentrated and makes up nearly a constant fraction  ($13-15\%$) of the gaseous atmosphere in the primary halo over the redshift range $z\lesssim 2.5$, regardless of whether the primary halo has formed or is still in the throes of active formation.  We also showed that a significant fraction of the gas that makes up the $z=0$ corona entered the halo via cold filamentary streams. It become part of the corona only after being heated to temperatures in access of $10^6$ K by SNe heating and merger-induced shocks, and expelled from the central galaxy.  

That mergers play an role in generating the warm-hot/hot medium around disk galaxies has been previously seen in idealized, non-cosmological simulations of merging galaxies \citep{Cox:2006, Sinha:2009}. The present study  confirms this also happens in galactic systems forming within a realistic cosmological setting. And, the impact of the merger-induced outflows is much more pronounced when supernova feedback is acting in concert.  With respect to the latter, we note that large-scale SNe-powered galactic winds are a crucial feature of the contemporary galaxy formation models, without which it is not possible to obtain realistic galaxy properties across the desired range of mass scales and galaxy types  \citep[e.g.][; see also \citealt{SomervilleDave:2015} and references therein]{In-N-OutBaryons:2016,Dave:2016,GK:2017,Illustris:2018}

%In the outskirts of the forming galactic halo, the early presence of the warm-hot atmosphere, regardless of whether it is in hydrostatic equilibrium or gently expanding, gives rises to back pressure that leads to the formation of accretion shocks earlier than predicted by models did not account for strong galactic wind 

The wind-fed corona and the winds themselves play a pivotal role in regulating flow of gas from the cosmic web to the galaxy.  In the absence of thermal winds, the filaments funnel cold gas onto the galaxy until the evolution of the large-scale structure causes the filaments to thicken and the gas accretion model to become diffuse and quasi-spherical.  We find that powerful galactic winds can disrupt the filament-galaxy connection well before the cosmological transition happens, repeatedly detaching the galaxy from its filamentary network and even accelerating the destruction of the connection altogether.  In other words, the winds and the corresponding corona take on a preventive role, in addition to their well-established ejective role.  This is an intriguing twist with profound implications.  For one, it would mean that all ``bathtub-type'' galaxy formation models, a category that includes most of the numerical simulations that use the decoupled kinetic winds prescription \citep[see][ for a full list of simulation studies that model galactic winds in this fashion]{SomervilleDave:2015}, are inadequate because they focus exclusively on the ejective aspect of the galactic winds.  In this class of models, galaxy evolution is governed primarily by the competition between gas accretion and expulsion, and the mass loss rate is tied to the star formation rate \citep{Bouche:2010,Lilly:2013}.  In light of our findings, we have initiated  a much more thorough analysis of galactic winds as ``agents'' of both ejective and preventative feedback.

Finally, apart from the above puzzle about the action of galactic winds, our study leaves unresolved several other issues.  For instance, given our small number of simulations and the restricted choice of sub-grid feedback models adopted, one could wonder how our key findings would change had we considered  a much larger sample of initial conditions as well as sub-grid recipes for feedback.  One thing is clear, though. The assembly history of the galaxy is an important factor especially since the corona appears to act in a preventive feedback capacity and mergers play a crucial role in generating the hot corona at early times by triggering both shock-heated outflows as well as starbursts that further boost the outflows.  It then follows that the number of major mergers occurring before the halo has grown enough to sustain a stable shock may be important in determining when a spiral galaxy begins to quench.  We may have already seen a hint of this within our simulation set: The growth of the hot corona in \venus, which has a much more active merging history, departs considerably from the predictions of the standard accretion shock picture when compared to \eris.  We speculate that for systems of comparable final mass, perhaps quenching is more effective and starts sooner in spiral galaxy systems with active merging history.   If borne out, this could be a major new addition to the current paradigm for galaxy formation and could, for example, help explain the scatter in the stellar-to-halo mass relationship in the vicinity of the Milky Way mass scale \citep[c.f.][]{Moster:2018}.  This is yet another aspect to be investigated.
}

\section{Acknowledgements}
Authors would like to thank Simon White, Andrey Kravtsov, Crystal Martin, Jerry Ostriker, Joop Schaye, Ali Rahmati, Sebastian Trujillo-Gomez, Mike Fall, Neal Katz and Carlos Frenk for valuable discussions that contributed to this paper. L.M. thanks the Kavli Institute for Theoretical Physics at UC Santa Barbara
for hospitality during the ``Galaxy-Halo Connection" Program in Spring 2017, during which preliminary results of this work were presented and discussed. AB acknowledges support from NSERC (Canada) through the Discovery Grant program and from the Pauli Center for Theoretical Studies ETH UZH, and is grateful to the University of Zurich's Institute for Computational Sciences, and especially the members of the Institute's Center for Theoretical Astrophysics and Cosmology, for  hospitality during his extended visits in Winter 2016 and Spring 2017.

\bibliographystyle{aasjournal}
\bibliography{references}
\end{document}